\documentclass[12pt]{article}
\usepackage[utf8]{inputenc}
\usepackage{amsmath,amsfonts,amsthm}
\usepackage{amssymb}
\usepackage{geometry}
\usepackage[superscript]{cite}
\usepackage{graphicx}
\usepackage{caption}
\usepackage{subcaption}
\usepackage{setspace}
\usepackage{lipsum}
\usepackage{mathtools}
\usepackage{cuted}
\usepackage[english]{babel}
\usepackage{hyperref}
\usepackage{multicol}
\usepackage[dvipsnames]{xcolor}
\usepackage{multirow}
\usepackage{multirow}
\usepackage{tabularx}
\usepackage{booktabs}
\usepackage{multicol}
\usepackage{mhchem}
\usepackage{rotating}
\usepackage{booktabs}
\usepackage{array}
\usepackage{makecell}
\usepackage{multirow}
\usepackage{pifont}
\usepackage{amssymb}
\usepackage{fontawesome}

\title{\textbf{Interplay of Altermagnetism and Coupled Quasi-Altermagnetic states in Sliding Two-dimensional Square Lattice}}
\author{Bhautik R Dhori$^{1}$, Deepak Upadhyay$^{1}$, and Prafulla K Jha$^{1,*}$}
\date{\small$^{1}$\textit{Department of Physics, Faculty of Science, The Maharaja Sayajirao University of Baroda, Vadodara, Gujarat 390002, India.}\\ \footnotesize Corresponding Authors: $^{*}$prafullaj@yahoo.com}

\begin{document}

\maketitle
\doublespacing

\begin{abstract}
	
\doublespacing

The emergence of non-relativistic spin splitting (NRSS) in altermagnetic systems has introduced a new paradigm in antiferromagnets with vanishing net magnetization. Although sliding-induced valley-polarized phases have recently been demonstrated in two-dimensional altermagnets, the observed valley-polarized state represents only a partial manifestation of altermagnetism, and a comprehensive classification based on spin-splitting characteristics remains lacking. Here, using \textit{first-principles calculations}, general stacking theory, and spin-Laue symmetry analysis, we propose a \textit{coupled quasi-altermagnetic state} representing a distinct subclass of altermagnetism, in which reversible type-IV NRSS is controlled through interlayer sliding. Accordingly, the sliding-induced phases are classified into two categories: altermagnetic and quasi-altermagnetic states. We establish a direct correspondence between reciprocal-space spin splitting and real-space switching between the two quasi-altermagnetic states. Importantly, the spin-polarized bands in these states remain spin split at $\Gamma$ point even in the absence of spin-orbit coupling (SOC), distinguishing them within the proposed classification framework. To demonstrate the interplay between altermagnetic and quasi-altermagnetic states, we investigate the two-dimensional Lieb-lattice material Mn$_2$WS$_4$ and its Janus derivative Mn$_2$WS$_2$Se$_2$, analysing how changes in the local environment influence the different magnetic phases. Importantly, the underlying mechanism is broadly applicable to a wide class of two-dimensional square-lattice systems. We further investigate the effects of SOC, focusing on spin texture and transport signatures in coupled quasi-altermagnetic states. Owing to the breaking of $\mathcal{C}_{4z}$ symmetry, which connects the sublattices in the tetragonal quasi-altermagnetic phase, anomalous Hall conductivity emerges and exhibits opposite-sign reversal behaviour. Our study uncovers a sliding-induced altermagnetic and quasi-altermagnetic phases and reveal their crucial role in controlling spin degeneracy at the Brillouin-zone centre, thereby paving the way for future spintronic technologies.
 
\end{abstract}
\
\textit{keywords} : Altermagnetism, Quasi-Altermagnetic State, Two-dimensional Material

\section{Introduction}
  Antiferromagnets (AFMs) have long been regarded as the paradigmatic example of magnetic order without net magnetization, where opposite spin sublattices cancel each other in both real and reciprocal space \cite{nvemec2018antiferromagnetic,wolf2001spintronics}. The preserved time-reversal ($\mathcal{T}$) symmetry combined with inversion or translation leading to largely spin-degenerate bands. Altermagnets (AMs) reshape this familiar picture by $\mathcal{T}$ symmetry breaking sizable spin-split bands across the Brillouin zone even without net magnetization and spin-orbit coupling (SOC) effects \cite{vsmejkal2022emerging,vsmejkal2022beyond,krempasky2024altermagnetic}. AMs unite a spin-split bands like ferromagnets (FMs) and hallmark traits of compensated sublattices of AFMs, making FM-AFM intersection by crystal's proper/improper rotational symmetry \cite{bai2024altermagnetism,song2025altermagnets,fender2025altermagnetism,jungwirth2026symmetry,zhou2025transition}. As a result, AMs enables efficient spin-polarized manipulate currents while retaining zero net magnetization, offering a unique advantage for spintronic applications. Moreover, spin-splitting in AMs is directly tied with the degree of spin valley polarization, making them ideal for valleytronics applications and can be used as information storage and carriers \cite{ding2024large,yang2025three}. A wide range of three-dimensional AMs has been theoretically proposed, including NiAs-type MnTe \cite{osumi2024observation,belashchenko2025giant} and CrSb \cite{reimers2024direct,yang2025three,zhou2025manipulation}, rutile RuO$_2$ \cite{liu2024absence,adamantopoulos2024spin,qian2026determining} and MnF$_2$ \cite{sadhukhan2026first,hariki2024determination}, perovskite SrRuO$_3$ \cite{naka2025altermagnetic} and LaMnO$_3$, and several other material classes. Encouragingly, non-relativistic spin splitting (NRSS) in $\alpha$-MnTe \cite{krempasky2024altermagnetic} and CrSb \cite{reimers2024direct} has been experimentally verified using soft X-ray angle-resolved photoemission spectroscopy.
  
  Spin-splitting in AMs can be classified into two distinct classes based on spin winding. First, altermagnetic quasiparticles exhibiting quadratic dispersion around high symmetry point $\Gamma$, characterized by spin-winding number 2. This significantly differs from the Rashba model (spin-winding number 1) \cite{vsmejkal2022emerging,turek2022altermagnetism}. Second, altermagnetic spin-polarized valley quasiparticle exhibit \textit{C-paired spin-valley locking (SVL)}, at time reversal invariant momenta (TRIM) $k$ and $-k$, without spin-winding \cite{zhang2025crystal,zhang2025strain,hu2025catalog}. Notably, this C-paired SVL in AM is fundamentally different from the T-paired SVL in transition-metal dichalcogenides (TMDs), which coupled spin and valley to real space through crystal symmetry and observed without SOC \cite{ma2021multifunctional,schaibley2016directional}. As a model, several two-dimensional (2D) monolayers V$_2$Se$_2$O \cite{qi2024spin,zeng2026crystal}, Cr$_2$O$_2$ \cite{liu2026uncompensated}, Cr$_2$SO \cite{guo2023piezoelectric} and Fe$_2$WTe$_4$ \cite{fan2025valley,li2025ferrovalley} have been predicted. The spin-valley manipulation accompanied by giant piezomagnetism has been observed in 2D V$_2$Se$_2$O crystal \cite{zhang2025crystal}. In addition, 2D CrO displays pronounced spin splitting, spin-momentum locking, and a high transition temperature. \cite{yang2025spin}. The weak interlayer van der Waals (vdW) coupling in 2D systems provide a versatile platform for tuning spin-valley properties via stacking engineering \cite{guo2024layer,guo2024electric,guo2025first}. Recently, coupling of spin-valley and layer configuration has led to the realization of altermagnetism in bilayer metal-organic framework systems such as Cr(tcb)$_2$ and Cr(hcb)$_2$ \cite{che2025bilayer}. The stacking engineering and magnetic order in bilayers are governed by $\mathcal{PT}$ symmetry, where $\mathcal{P}$ denotes spatial inversion symmetry. In bilayer systems, the layer-dependent out of plane pseudospin component corresponds to an electric dipole moment, in contrast to monolayers where only in-plane wave-vector contributions exist. This dipole moment can be tuned using a perpendicular electric field E$_z$.  
  
  Moreover, extensive theoretical studies indicate that spin-valley properties in 2D AMs can be engineered through electric fields, strain, stacking, and interlayer sliding \cite{wang2024electric,zhang2025strain,xun2025stacking,cheng2026sliding}. An effective route to achieve valley polarization (VP) in AMs is through coupling between the AM ground state and ferroelectricity giving rise to the altermagnetoelectric effect. A key advantage of this coupling is the ability to tune altermagnetic spin splitting via ferroelectric switching (i.e., spontaneous polarization P $\uparrow$ $\leftrightarrow$ P $\downarrow$), enabling a potential pathway for electrically driven non-volatile switching in AM systems \cite{gu2025ferroelectric,guo2025mechanically}. In vdW bilayer configurations, out-of-plane polarization emerges from inversion symmetry breaking induced by interlayer translation. Consequently, switching between distinct polar stacking configurations (AB $\leftrightarrow$ BA) reverses P, thereby reversing the altermagnetic spin splitting. This mechanism enables non-volatile modulation of layer-sliding-induced VP in altermagnetic-ferroelectric(ferroelastic) systems. As a model system, 2D monolayers MnPTe$_3$ \cite{sun2025designing}, RuF$_4$ \cite{peng2025ferroelastic}, and CuF$_2$ \cite{peng2026sliding} have been predicted. Previous studies have identified sliding as an efficient route to achieve VP in altermagnetic bilayers spanning diverse 2D materials. Nevertheless, a systematic exploration, detailed categorization of distinct spin-splitting phenomena in altermagnets, and a comprehensive theoretical framework for the coupling between different states remain absent.
  
  In this paper, we investigate ``coupled quasi-altermagnetic states'', representing a distinct subclass of altermagnetism in which reversible type-IV NRSS is modulated through interlayer sliding. We demonstrate that the previously reported sliding-induced valley-polarized states in altermagnetism provide an incomplete description of the non-relativistic spin-splitting. Accordingly, we classify sliding-induced phases into two categories: altermagnetic states and quasi-altermagnetic states. Detailed symmetry analysis, together with an examination of spin degeneracy at the $\Gamma$-point, reveals that these phases cannot be assigned to the same class. Furthermore, we reveal a direct correspondence between spin splitting in reciprocal space and real-space magnetic configuration switching between the two quasi-altermagnetic states. Importantly, analogous to ferroic mechanisms in altermagnetic systems, including ferroelectricity and ferroelasticity, the proposed coupled quasi-altermagnetic states also lead to reversible sign changes in transport properties such as the anomalous Hall conductivity (AHC). Our analysis provides in-depth insight into quasi-altermagnetic states and their consequences at the $\Gamma$-point, while also introducing a classification framework for sliding-induced phases.
  
\section{Methodology}

All first-principles calculations were carried out within the framework of density functional theory as implemented in the Vienna ab initio simulation package (VASP) \cite{kresse1994ab} and the projector augmented-wave (PAW) method \cite{blochl1994projector,kresse1999ultrasoft}. The exchange-correlation functionals were treated by the generalized gradient approximation (GGA) with the Perdew-Burke-Ernzerhof (PBE) parametrization \cite{perdew1996generalized}. A vacuum layer of 30 $\AA$ was included along the vertical direction to suppress interactions between periodic images. Structural optimizations were performed using a plane-wave kinetic energy cutoff of 600 eV, force convergence criterion of 0.001 eV/$\AA$, and total energy convergence criterion of $10^{-7}$ eV. The Brillouin zone (BZ) was sampled using a 10$\times$10$\times$1 Monkhorst-Pack k-point mesh \cite{38}. To accurately describe the strongly correlated Mn-3d electrons, the rotationally invariant DFT+U approach was employed with an on-site Hubbard parameter of U = 3.0 eV. This value is consistent with previous studies on Mn$_2$WS$_2$ and reliably reproduces the electronic band gap and magnetic moments. For all investigated stacking configurations and magnetic orderings, the total magnetic moment per unit cell remained negligible $<$ $10^{-3}$ $\mu_B$ , indicating fully compensated magnetic states characteristic of antiferromagnetic behaviour. Following the convergence of the ground-state charge density, the tight-binding Hamiltonian was constructed using maximally localized Wannier functions as implemented in the Wannier90 code \cite{mostofi2008wannier90,pizzi2020wannier90}. Spin-orbit coupling was included in the AHC calculations, which were performed using the Kubo formalism by integrating the Berry curvature over all occupied bands in momentum space. The AHC is given by

\begin{equation}
\sigma_{\alpha\beta}^{\mathrm{AH}}
=
\frac{e^2}{\hbar}
\int_{BZ}\frac{d\mathbf{k}}{(2\pi)^3}
\,\Omega_{\alpha\beta}(\mathbf{k})
\end{equation}
Where $\alpha,\beta$ denotes the cartesian direction, $\Omega_{\alpha\beta}(\mathbf{k})$  is Berry curvature, which can be expressed as
\begin{equation}
\Omega_{\alpha\beta}(\mathbf{k})
=
\sum_n f_{kn}\,\Omega_{n,\alpha\beta}(\mathbf{k})
\end{equation}
where, $f_{kn}$ is Fermi-Dirac distribution function and $\Omega_{n,\alpha\beta}(\mathbf{k})$ is band-projected Berry curvature-like term

\begin{equation}
\Omega_{n,\alpha\beta}(\mathbf{k})
=
2i
\sum_{m\ne n}
\frac{
	\langle n|\partial \hat{H}/\partial k_\alpha|m\rangle
	\langle m|\partial \hat{H}/\partial k_\beta|n\rangle
}{
	\left(\epsilon_n^k-\epsilon_m^k\right)^2
}
\end{equation}
where $\partial \hat{H}/\partial k_{\alpha, \beta}$ is velocity operators, while $|n\rangle$ and $\epsilon_n^k$ represent the eigenvectors and eigenvalues of the $n^{th}$ band at momentum $k$. Crystallographic and symmetry analyses were further performed using the Bilbao Crystallographic Server \cite{aroyo2006bilbao}, FINDSYM \cite{stokes2005findsym}, and the approach described in Ref. \cite{pan2024general,zhu2025sliding}. In addition, several auxiliary packages and databases, including VASPKIT \cite{wang2021vaspkit}, VESTA \cite{momma2008vesta}, Materials Project \cite{jain2013commentary}, PyProcar \cite{herath2020pyprocar}, and MAGNDATA \cite{gallego2016magndata}, were utilized for structural visualization, data post-processing, and electronic structure characterization.

\section{Results and Discussion}

\subsection{Symmetry Considerations}
AMs exhibit SOC-independent spin splitting and are therefore naturally characterized by spin-group symmetries $[R_i||R_j]$, where $R_i$ and $R_j$ acting independently govern spin space and real space, respectively. Collinear magnets are characterized by a spin-only allowed symmetry operations are $R_i \in \{E\}\ \text{or}\ \{E,\overline{\mathcal{C}_2}\}$, where $E$ and $\overline{\mathcal{C}_2}$ represents an identity operation and twofold rotation about an axis perpendicular to spin alignment followed by spin inversion, respectively. Under this operation, the energy eigenvalues transform as $\varepsilon(s,\mathbf{k}) \rightarrow \varepsilon(s,-\mathbf{k})$, giving rise to even-parity spin splitting. However, when combined symmetries such as $[\overline{\mathcal{C}_2} \| \mathcal{T}][\mathcal{C}_2 \| \mathcal{P}]$ and  $[\mathcal{C}_2 \| \mathcal{\tau}]$ are present, the transformation becomes $\varepsilon(s,\mathbf{k}) \rightarrow \varepsilon(-s,\mathbf{k})$, which restores spin degeneracy at any $k$ point. In general, altermagnetic spin-splitting represented by a non-trivial spin group $R_s = [E||H]+[\mathcal{C}_2||G-H] = [E||H]+[\mathcal{C}_2||AH]$, where $H$ is halving subgroup of space group $G$ and $A\in G-H$. In contrast to 3D case, where $A\in \{\tau,\mathcal{P}\}$, 2D representation imposes a constrain $A \notin \{\tau, \mathcal{P}, \mathcal{C}_{2z}, \mathcal{M}_{z}\}$, within condition that 2D collinear magnet restricted in $x-y$ plane. For example, in 2D case, two spin group symmetries $[\mathcal{C}_2 \| \mathcal{M}_{z}]$ and $[\mathcal{C}_2 \| \mathcal{C}_{2z}]$ transform as $\varepsilon(s,\mathbf{k}) \rightarrow \varepsilon(s,-\mathbf{k})$, protect Kramers spin degeneracy. In $[\mathcal{C}_2 \| \mathcal{M}_{z}]$, opposite spin sublattices can be connected through $\mathcal{M}_{z}$ symmetry through $x-y$ plane, $[\mathcal{C}_2 \| \mathcal{M}_{z}]\varepsilon(s,\mathbf{k})=\varepsilon(-s,\mathbf{k})$. Although, $[\mathcal{C}_2 \| \mathcal{M}_{z}]$ symmetry also exists in the system, $[\mathcal{C}_2 \| \mathcal{M}_{z}]\varepsilon(s,\mathbf{k})=\varepsilon(s,\mathbf{k})$, which ensures spin degeneracy \cite{zeng2024description}. The second symmetry operation $[\mathcal{C}_2 \| \mathcal{C}_{2z}]$, opposite spin sublattices can be connected through two-fold rotational symmetry through z axis. The product of $[\mathcal{C}_2 \| \mathcal{C}_{2z}]$$[\overline{\mathcal{C}_2} \| \mathcal{T}]$ gives rise to $\varepsilon(s,\mathbf{k}) \rightarrow \varepsilon(-s,\mathbf{k})$, which restores spin degeneracy in 2D system. Accordingly, these four symmetries $\{\tau, \mathcal{P}, \mathcal{C}_{2z}, \mathcal{M}_{z}\}$ must be excluded in the observation of a non-trivial spin layer group associated with altermagnetism.  Nevertheless, the presence of at least one symmetry relating these sublattices is essential to guarantee zero net magnetization enforced by symmetry operations like in-plane two-fold rotation $[\mathcal{C}_2 \| \mathcal{C}_{2x}]$, out of plane four-fold rotation $[\mathcal{C}_2 \| \mathcal{C}_{4z}]$, and vertical mirror plane $[\mathcal{C}_2 \| \mathcal{M}_{x}]$ in 2D altermagnets. 

\begin{figure}[ht!]
	\centering
	\includegraphics[width = 15cm]{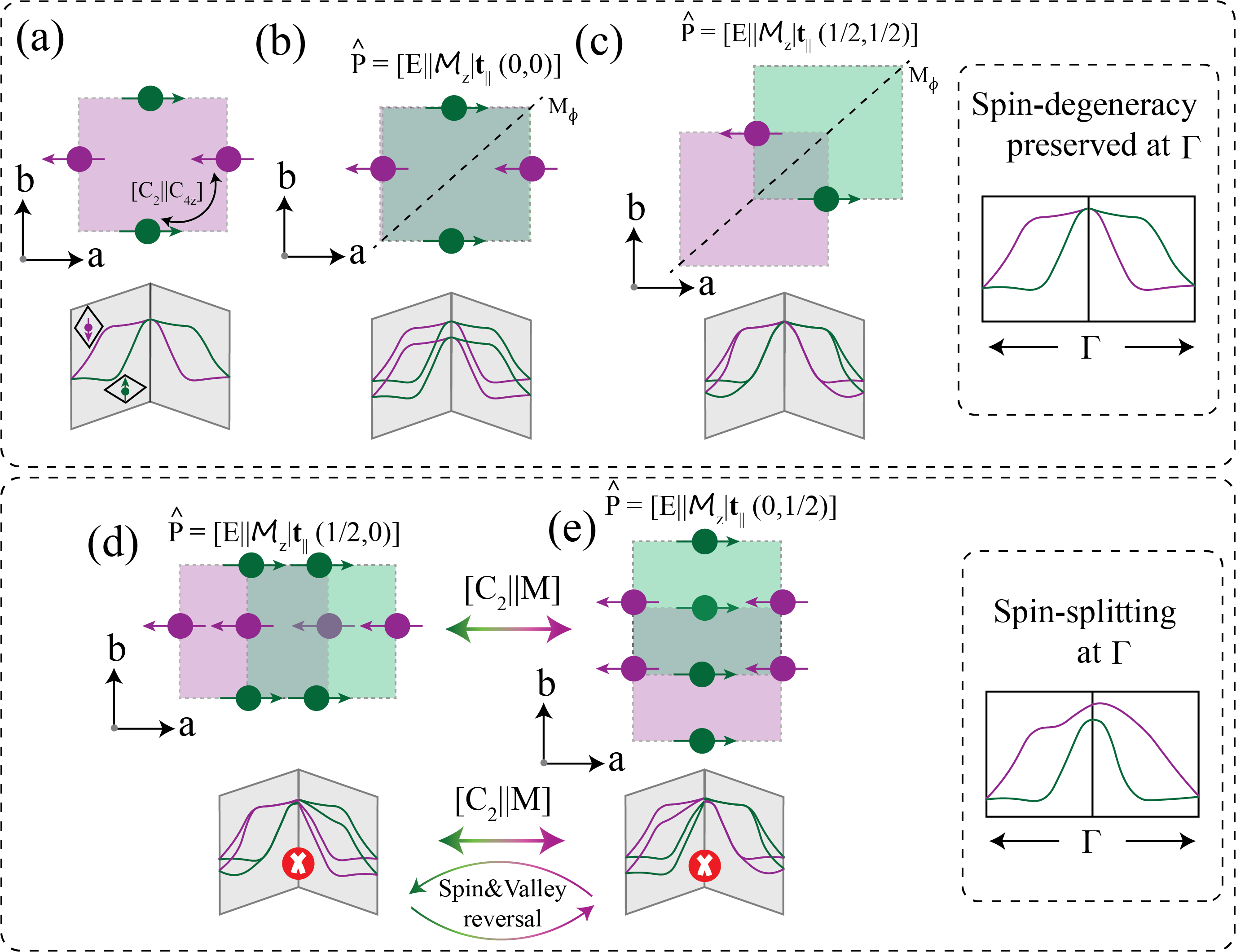}
	\caption{\label{sche}Schematic illustration of the two-dimensional square-lattice model. The $\uparrow$ and $\downarrow$ sublattices are represented by green and purple circles, respectively. The monolayer $L$ is denoted by the pink square, while its counterpart $L^{\prime}$ is shown in green. The configurations are classified into two categories according to whether spin degeneracy at the $\Gamma$ point is preserved or broken. (a) Monolayer configuration, illustrating d-wave altermagnetism arising from the underlying combined $[\mathcal{C}_2 \| \mathcal{C}_{4z}]$ symmetry operations.(b) AA-bilayer configuration generated by the symmetry operation $\widehat{P}=[E \|\mathcal{M}_{z}|t_{\parallel}(0,0)]$, in which the second monolayer lies directly above the first (c) AB-bilayer configuration generated by the symmetry operation $\widehat{P}=
		\left[E \| \mathcal{M}_{z}| t_{\parallel}
		\left(\frac{1}{2},\frac{1}{2}\right)\right]$ where the second monolayer is slide along the [110] direction. For both configurations with $t_a=t_b$, the $[\mathcal{C}_2 \| \mathcal{M}_{\phi}]$ symmetry is preserved and a valid opposite-spin sublattice connector exists.(d) AC$_1$  configuration and (e) AC$_2$ configuration, generated by the stacking operations $\widehat{P}=
		\left[E \| \mathcal{M}_{z}| t_{\parallel}
		\left(\frac{1}{2},0 \right)\right]$ and $\widehat{P}=
		\left[E \| \mathcal{M}_{z}| t_{\parallel}
		\left(0,\frac{1}{2}\right)\right]$, respectively. The $G–H$ set is empty in both configurations, while spin splitting is preserved throughout BZ. The AC$_1$ and AC$_2$ bilayer structures belong to the quasi-altermagnetic phase. Two states are connected through the real-space symmetry operation $[C_2 \| M]$, which induces simultaneous spin and valley reversal, thereby forming a pair of coupled quasi-altermagnetic states.}
\end{figure}

We begin with a 2D square d-wave AM lattice model as the basic building unit. This system comprises a opposite sublattices having zero net magnetization protected by $[\mathcal{C}_2 \| \mathcal{C}_{4z}]$ rotational symmetry (Fig.\ref{sche}(a)). Importantly, the in-plane diagonal mirror symmetry $\mathcal{M}_{\phi}$, with mirror plane oriented perpendicular to the $[1\bar 10]$ direction, linked to the $X$ and $X^{\prime}$ valley. Furthermore, $\Gamma$-point is invariant under $[\mathcal{C}_2 \| \mathcal{C}_{4z}]$ real space transformations enforce spin degeneracy at $\Gamma$. To construct the bilayer, two AM monolayers $L$ and its counterpart $L^{\prime}$ were stacked using a designated stacking operator $\widehat{P}$, where  $\widehat{P}=\{O|t_{0}\}$. Here, $O$ and $t_{0}$ are rotational and translational operator, respectively. The counterpart $L^{\prime}$ is generated through $\widehat{P}$ as $\widehat{P}L$, which leads to the $L+\widehat{P}L$ bilayer configuration. Distinct choices of $\widehat{P}$ applied to the same layer  yield bilayers with different symmetries. We assume that two bilayers $BL_1 = L + \{O|t_{01}\}L
\quad \text{and} \quad
BL_2 = L + \{O|t_{02}\}L$ are generated through the same rotation $O$, while different in their translation vectors $t_{01}$ and $t_{02}$. We use the bilayer coordinate system, where the monolayers $L$ and $L^{\prime}$ are symmetrically arranged below and above the $k_z = 0$ plane, respectively. 

In this work, the stacking operation $\widehat{P} = [E \| \mathcal{M}_{z}|t_{\|}]$ is considered, where $t_{\parallel}=t_a\vec{a}+t_b\vec{b}$ denotes translation vector parallel to the layer. Hereafter, $t_{\parallel}=t_a\vec{a}+t_b\vec{b}$ will be denoted as $t_{\parallel}(t_a,t_b)$. Beginning with the AM, AA-bilayer configuration depicted in Fig.\ref{sche}(b) is constructed using $\widehat{P}=[E \|\mathcal{M}_{z}|t_{\parallel}(0,0)]$, without disturbing the in-plane atomic arrangement, which breaks $[\mathcal{C}_2 \| \mathcal{P}]$ and $[\mathcal{C}_2 \| \mathcal{M}_{z}]$ symmetries. Consequently, the observed spin splitting in the energy bands indicates that altermagnetism is preserved in the system. In the AA configuration, the $[\mathcal{C}_2 \| \mathcal{M}_{\phi}]$ symmetry operation relating the wave vectors $k_x$ and $k_{x^{\prime}}$ ensures the preservation of altermagnetic characteristics such that $E_{\nu}^{c}(k_x)=E_{\nu}^{c}(k_x^{\prime})$. The AB bilayer configuration (Fig.\ref{sche}(c)), constructed using $\widehat{P}=
\left[E \| \mathcal{M}_{z}| t_{\parallel}
\left(\frac{1}{2},\frac{1}{2}\right)\right]$, retains $[\mathcal{C}_2 \| \mathcal{M}_{\phi}]$ symmetry and consequently exhibits the same degeneracy. Notably, the AB stacking differs from AA in that the spin splitting between equivalent spin channels at the $X$ and $X^{\prime}$ valleys is suppressed. Accordingly, the monolayer, AA- and AB-bilayer configurations preserve altermagnetic order and spin degeneracy at $\Gamma$ high symmetry point remain intact without SOC. 

Moreover, AC$_1$ and AC$_2$ configurations (Fig.\ref{sche}(d,e)) are constructed using $\widehat{P}=
\left[E \| \mathcal{M}_{z}| t_{\parallel}
\left(\frac{1}{2},0 \right)\right]$ and $\widehat{P}=
\left[E \| \mathcal{M}_{z}| t_{\parallel}
\left(0,\frac{1}{2}\right)\right]$, respectively, by translating one layer relative to the other along $\left(\frac{1}{2},0 \right)$ and $\left(0,\frac{1}{2} \right)$ directions, respectively, while keeping the second layer fixed. This interlayer displacement breaks the $\mathcal{M}_{\phi}$ symmetry and still retain spin-splitting in BZ. However, these configurations do not satisfy the defining condition of altermagnetism, namely the absence of mapping between opposite-spin sublattices. For the AC$_1$ and AC$_2$ configurations, band degeneracy at $\Gamma$-point is lifted without inclusion of SOC. This stands in contrast to the three configurations (monolayer, AA- and AB-bilayer) of altermagnetic systems discussed earlier, where $\Gamma$-point states are strictly degenerate. This phase can be classified as a distinct subclass of spin-splitting type-IV compensated antiferromagnets, known as \textit{``quasi-altermagnet''}. The quasi-AM phase lacks $\mathcal{PT}$ symmetry, a valid spin-exchanging connector, and $A \notin \{\tau, \mathcal{P}, \mathcal{C}_{2z}, \mathcal{M}_{z}\}$ degeneracy-enforcing set.  Therefore, opposite-spin sublattices are not connected by rotational symmetry, yielding spin splitting and fully lifted spin degeneracy, even at $\Gamma$. Quasi-altermagnets can be understood as systems derived from altermagnets through rotational symmetry breaking, resulting in the empty $G-H$ set. They exhibit anisotropic momentum-dependent spin splitting that is not equal and opposite across the nodal plane (Fig.\ref{sche}(d,e)). Furthermore, the absence of a well-defined nodal plane leads to complete lifting of spin degeneracy while preserving zero net magnetization in insulating systems. Consequently, quasi-altermagnets may be classified as a subclass of altermagnetic systems. Fig.\ref{sche}(d,e) illustrates two quasi-altermagnetic states, $a$ (AC$_1$ configuration) and $b$ (AC$_2$ configuration), exhibit complete compensation of opposite spin sublattices and remain invariant under transformation between one another, in the absence of any ferroelectric order. The spin sublattice configuration of quasi-altermagnetic state $a$ is related to that of state $b$  through an orthogonal transformation $O_s$, which leads to $\varepsilon_b(s,\mathbf{k})
=
[C_2 \parallel O_s]\,
\varepsilon_a(s,\mathbf{k})
=
\varepsilon_a(-s,O_s\mathbf{k})$, where $O_s$ can be mirror reflection or rotation, followed by without or with translation \cite{chen2025cross}. The formulation reveals that quasi-altermagnetic states $a$ and $b$ are related via a $[C_2 \| O_s]$ real-space transformation, leading to spin-valley reversal signature. In particular, the spin-up (spin-down) band of state $a$ at the $X(X^{\prime})$ valley maps onto the spin-down (spin-up) band of state $b$ at $X^{\prime}(X)$. These states can therefore be classified as \textit{``coupled quasi-altermagnetic''} states with zero net magnetization. While they do not strictly fulfil the criteria for altermagnets, they display finite spin splitting and ferromagnet-like switch ability. The $t_{\parallel}(t_a,t_b)$, where $t_a \neq t_b$ sliding plays a dual role by inducing these coupled quasi-altermagnetic states and spin-splitting at $\Gamma$ point. However, the coupled quasi-altermagnetic configuration is governed not only by the magnetic ordering but also significantly influenced by the surrounding non-magnetic atoms.

\subsection{Material Considerations}

\begin{figure}[ht!]
	\centering
	\includegraphics[width = 15cm]{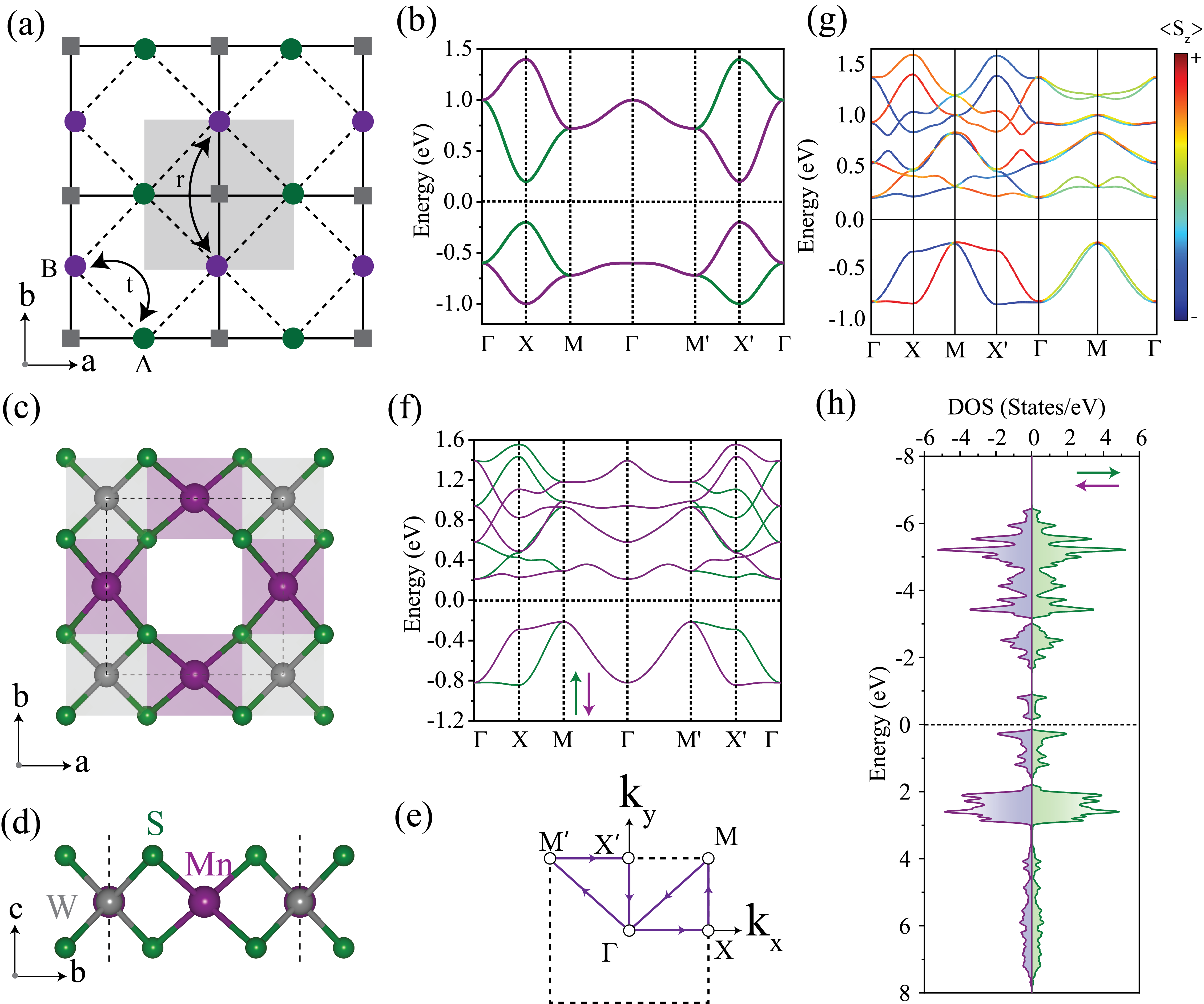}
	\caption{\label{material}(a) Schematic representation of the Lieb lattice in real space. Sublattices carrying opposite magnetic moments are represented by purple and green circles, while the non-magnetic site is denoted by a gray square. The spin-up and down states occupy the $2c$ Wyckoff positions, (0, 1/2) and (1/2, 0), respectively, whereas the $1b$ Wyckoff position hosts a non-magnetic atom. (b) Electronic band structure of the corresponding lattice model (c,d) Top and side views of Mn$_2$WS$_4$ structure. The Mn- and W-centered tetrahedra are highlighted in pink and gray, respectively. (e) BZ of square lattice indicating high-symmetry paths $\Gamma$-X-M and $\Gamma$-X$^{\prime}$-M$^{\prime}$. (f) Spin-polarized electronic band structure of monolayer Mn$_2$WS$_4$. (g) $S_z$-projected band structure with SOC. (h) Spin resolved DOS of Mn$_2$WS$_4$.
	 }
\end{figure}

To realize AM in square lattice, we construct a minimal 2D Lieb altermagnetic model comprising two magnetic sublattices carrying opposite magnetic moments (indicated by purple and green dots) and one non-magnetic site (Gray square), connected through four-fold rotational symmetry (Fig.\ref{material}(a)). The underlying magnetic symmetry is described by the space group ${P4'}/{mm'm}$ , with spin up and down states residing at $2c$ Wyckoff site $\left(0,\frac{1}{2}\right)$ and $\left(\frac{1}{2},0 \right)$, respectively, while the $1b$ site hosts a non-magnetic atom. Such a magnetic arrangement explicitly breaks $\mathcal{T}$ symmetry in space group ${P4'}/{mm'm}$ \cite{che2025engineering,durrnagel2025altermagnetic}. Using, $|d_{z^2};\uparrow\rangle$ and $|d_{z^2};\downarrow\rangle$ as basis functions, we include on-site energies $e$, nearest neighbour hopping $t$, and anisotropic next nearest neighbour hopping $r$, characterized by the parameters $e_1,\ e_2,\ t_1,\ t_2$ and $r_1-r_4$. The resulting Hamiltonian is a $4 \times 4$ matrix composed of $2 \times 2$ diagonal and off-diagonal blocks $H_{\mathrm{Lieb}}(k_x,k_y)
=
\begin{bmatrix}
	E_{11} & T_{12} \\
	T_{21} & E_{22}
\end{bmatrix}$. Detailed form of diagonal and off-diagonal blocks are given in supplementary material. Resulting band structure exhibiting altermagnetic characteristics, see Fig.\ref{material}(b), along $\Gamma(0,0)
\rightarrow
X(\pi,0)
\rightarrow
M(\pi,\pi)
\rightarrow
\Gamma(0,0)
\rightarrow
M'(-\pi,\pi)
\rightarrow
X'(0,\pi)
\rightarrow
\Gamma(0,0)$ high symmetry path. The $H_{\mathrm{Lieb}}$ is centrosymmetric along $\Gamma- M$ and $\Gamma - M^{\prime}$ paths, resulting in degenerate spin bands. Conversely, along $\Gamma \rightarrow X \rightarrow M$ and $\Gamma \rightarrow M^{\prime} \rightarrow X^{\prime}$ path, $H_{\mathrm{Lieb}}$ become non-centrosymmetric, leading to a maximum spin splitting of $\Delta E_{X(X^{\prime})}
= 4\left(r_{1(2)}-r_{3(4)}\right)$ at the $X(X^{\prime})$ point, classified as a d-wave AMs. 

Based on the insight obtained from the model Hamiltonian analysis and inspired by the family of 2D Lieb-lattice materials reported by Li Yang et al. \cite{xu2025alterpiezoresponse,xu2026axial,wang2026valley}, we considered monolayer Mn$_2$WS$_4$ (see Fig.\ref{material}(c)) as a candidate system for coupled quasi altermagnetism. The monolayer Mn$_2$WS$_4$ has previously been identified as a promising platform for altermagnetism. Structurally, it adopts a symmorphic tetragonal crystal structure with space group $P\bar{4}2m (No. 111)$, characterized by $D_{2d}$ point-group symmetry. Structurally, monolayer consists central Mn-W layer sandwiched between two S layers, forming a tetrahedral lattice geometry, see Fig.\ref{material}(d). Band structure of Mn$_2$WS$_4$ closely resembles the characteristic features of the 2D Lieb lattice model, particularly the altermagnetic spin splitting observed along the non-centrosymmetric paths $\Gamma \rightarrow X \rightarrow M$ and $\Gamma \rightarrow M' \rightarrow X'$ path, which is consistent with the previously reported results \cite{xu2025alterpiezoresponse} (see Fig.\ref{material}(f)). The resulting d-wave altermagnetism arises from the underlying combined $[\mathcal{C}_2 \| \mathcal{C}_{4z}]$ symmetry operations. In addition, the opposite spin sublattices and the $X$ and $X^{\prime}$ valleys are connected through $[\mathcal{C}_2 \| \mathcal{M}_{\phi}]$ spin symmetry. The BZ with high symmetry point is depicted in Fig.\ref{material}(e). In the SOC-free limit, symmetry protection arising from $[\mathcal{C}_2 \mathcal{T}\| \mathcal{C}_{2z}\mathcal{T}]$ constrains Berry curvature to vanish throughout BZ. Upon incorporating SOC, the spin-group symmetry is reduced to the magnetic-group symmetry, giving rise to finite Berry curvatures that are opposite in sign at $X$ and $X^{\prime}$ valleys. Notably, opposite spin polarization at the two valleys remains preserved, retaining characteristics similar to the non-relativistic band structure. These features give rise to a valley Hall effect (VHE) in Mn$_2$WS$_4$ subjected to an applied in-plane electric field. Importantly, with SOC, the combined $\mathcal{T}\mathcal{C}_{2z}$ is preserved within the condition that N\'eel vector lying within $ab$ plane. Consequently, the valley degeneracy is protected provided that mirror symmetry  is preserved.  However, SOC induces a pronounced spin splitting along the   direction, which becomes especially significant near the M point, see Fig.\ref{material}(g). Moreover, density of states (DOS) in the non-relativistic regime demonstrates complete compensation between the opposite spin sublattices (Fig.\ref{material}(h)).

\begin{figure}[ht!]
	\centering
	\includegraphics[width = 16cm]{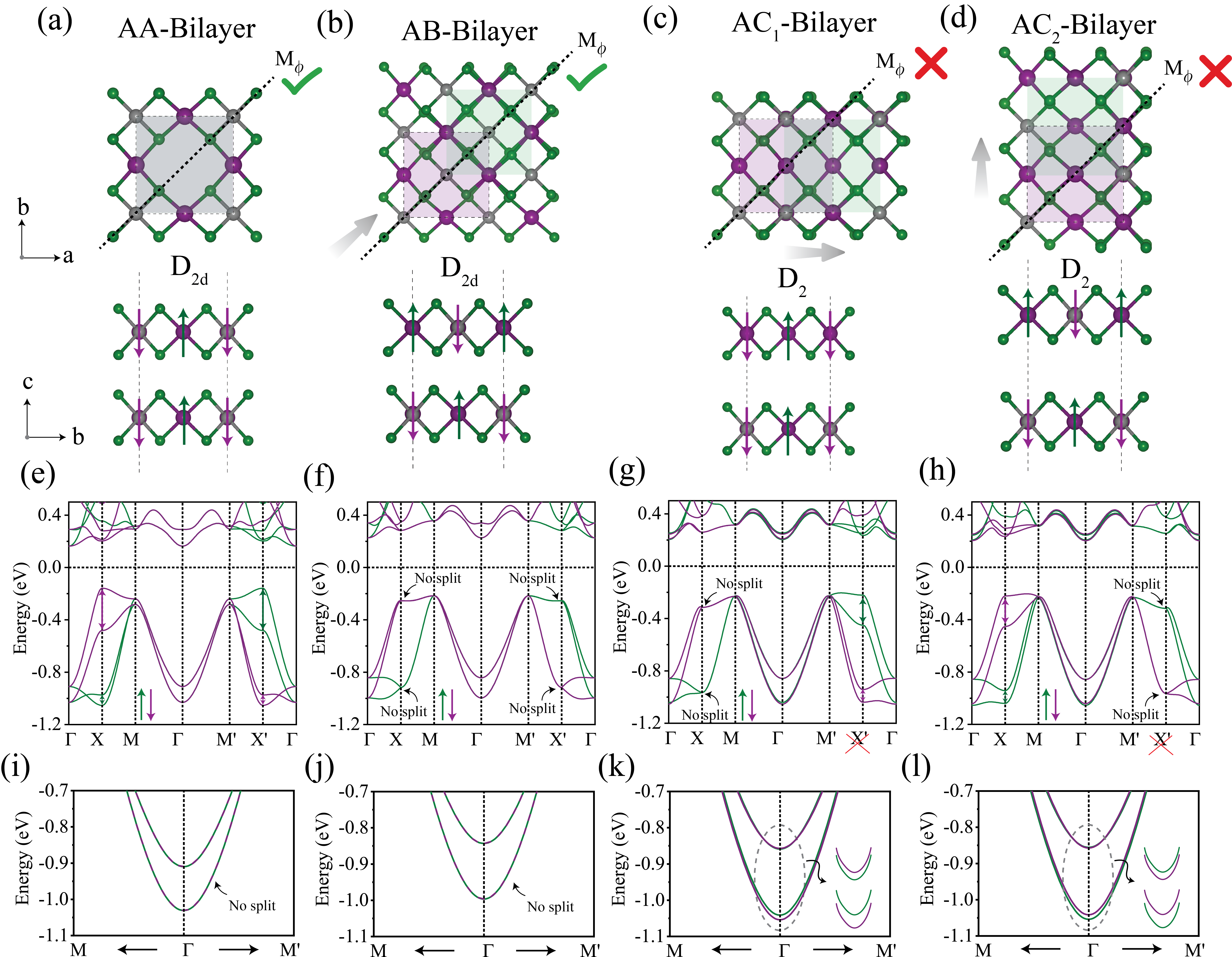}
	\caption{\label{bilayer}Crystal structures, spin-polarized band structures, and electronic band dispersions around the $\Gamma$ point for various bilayer configurations. Top and side views of crystal structures of (a) AA, (b) AB, (c) AC$_1$, and (d) AC$_2$ bilayer configurations. The preservation or breaking of the $\mathcal{M}_{\phi}$ symmetry in each configuration is indicated by green check marks and red crosses, respectively. The color scheme used for the monolayer and bilayer structures is consistent with that in Fig. 1. Green and purple arrows denote the $\uparrow$ and $\downarrow$ orientations of Mn atoms, respectively. The gray arrow denotes the direction along which the layer is translated during the sliding process. (e-h) Spin-polarized band structures corresponding to the AA, AB, AC$_1$, and AC$_2$ bilayer configurations, respectively. The red cross on the X$^{\prime}$ point indicates that the X and X$^{\prime}$ points are inequivalent and are not related by a rotational symmetry operation. Spin splitting at X and X$^{\prime}$ points within same spin channel is highlighted by the corresponding arrows.(i-j) Band structures along M-$\Gamma$-M$^{\prime}$ high-symmetry path. The insets show enlarged views of the bands near the $\Gamma$ point. Spin-degenerate bands are indicated by overlapping green and purple lines and labeled as No Split, whereas spin-split bands exhibit a finite energy separation. }
\end{figure}

The bilayer configurations are divided into two groups based on their altermagnetic characteristics. The AA- and AB-stacked bilayers strictly preserve altermagnetism and maintain spin splitting at the $\Gamma$-point. In contrast, the AC$_1$- and AC$_2$-bilayer configurations do not fully satisfy the altermagnetic condition and exhibit degeneracy lifting at the $\Gamma$-point. In first group, we constructed bilayer Mn$_2$WS$_4$ structures using $\widehat{P}=[E \|\mathcal{M}_{z}|t_{\parallel}(t_a,t_b)]$, where $t_a = t_b$, in the absence of $[\mathcal{C}_2 \| \mathcal{P}]$ and $[\mathcal{C}_2 \| \mathcal{M}_{z}]$ symmetries to assess the influence of interlayer sliding configuration on altermagnetic spin-splitting, see Fig.\ref{sche}. AA-bilayer configuration form by the second monolayer is positioned directly above to the first through $\widehat{P}=[E \|\mathcal{M}_{z}|t_{\parallel}(0,0)]$, where in plane structural geometry remain identical in both layers, without altering the in-plane atomic arrangement (see Fig.\ref{bilayer}(a)). The AA-Mn$_2$WS$_4$ configuration is characterized by the $P\bar{4}2m$ (spin space group: $P^{\bar 1}\bar{4}^{1}2^{\bar 1}m^{\infty 100m}1$) space group, with symmetry operations classified into the sets $H \in \{E,\, \mathcal{P},\, \mathcal{C}_{2z},\, \mathcal{M}_{z}\}$ and $G-H \in \{\mathcal{M}_{\varphi},\, \mathcal{M}_{\varphi'},\, \mathcal{C}_{2\varphi},\, \mathcal{C}_{2\varphi'}\}$. Consequently, AA-configuration exhibits altermagnetic behaviour exhibit alternating spin splitting and spin degeneracy at $\Gamma$, see Fig.\ref{bilayer}(e,i). 

\begin{figure}[ht!]
	\centering
	\includegraphics[width = 14cm]{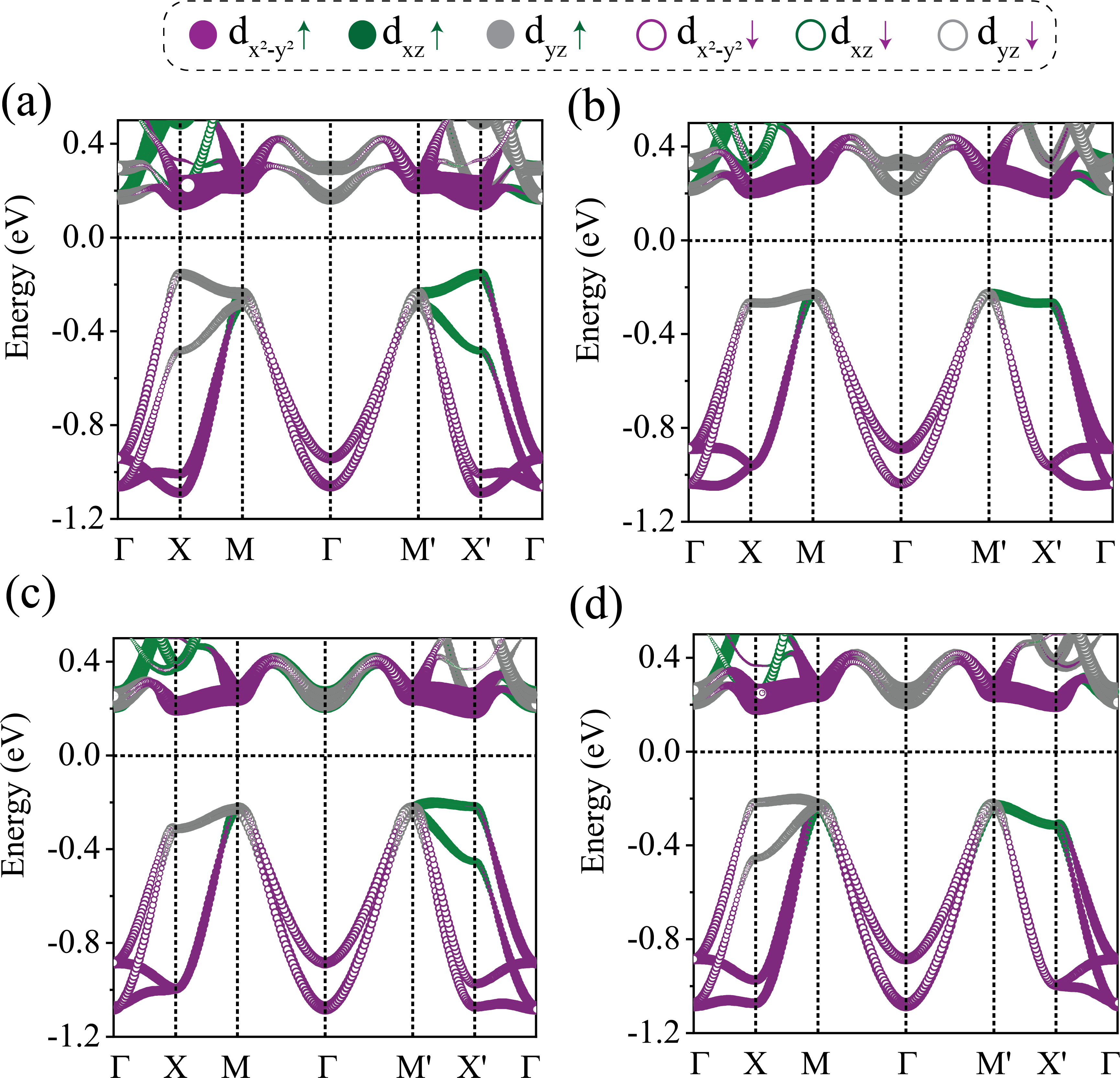}
	\caption{\label{bilayer_orbital}Mn $d$-orbital-projected band structures of the (a) AA, (b) AB, (c) AC$_1$, and (d) AC$_2$ bilayer configurations. The $d_{x^2-y^2}$, $d_{xz}$, and $d_{yz}$ orbital contributions are represented by purple, green, and gray circles, respectively, whose sizes are proportional to their corresponding projection weights. Filled and open circles denote $\uparrow$ and $\downarrow$ states, respectively. For a meaningful comparison of orbital contributions among different configurations, an identical projection-weight scaling factor is employed in all panels.}
\end{figure}

The AB-bilayer configuration is obtained from AA-bilayer by sliding one layer relative to other along the fractional coordinate $\left(\frac{1}{2},\frac{1}{2}\right)$, using $\widehat{P}=[E \|\mathcal{M}_{z}|t_{\parallel}\left(\frac{1}{2},\frac{1}{2}\right)]$ see Fig.\ref{bilayer}(b). This transformation drives Mn$_2$WS$_4$ system from $P\bar{4}2m$ (spin space group: $P^{\bar 1}\bar{4}^{1}2^{\bar 1}m^{\infty 100m}1$) space group to $P\bar{4}2_1m$ (spin space group: $P^{ 1}\bar{4}^{1}2^{1}_{1}m^{\infty 100m}1$) space group, with symmetry operations grouped into the sets $H \in
\left\{
E,\,
\mathcal{C}_{2z},\,
\left[\mathcal{P}\,\big|\,t_{\parallel}\left(\frac{1}{2},\frac{1}{2}\right)\right],\,
\left[\mathcal{M}_{z}\,\big|\,t_{\parallel}\left(\frac{1}{2},\frac{1}{2}\right)\right]
\right\}$ and $G-H \in
\left\{
\mathcal{M}_{\varphi},\;\mathcal{M}_{\varphi^{\prime}},\;
\left[\mathcal{C}_{2\varphi}\,|\,t_{\parallel}\left(\frac{1}{2},\frac{1}{2}\right)\right],\;\left[\mathcal{C}_{2\varphi^{\prime}}\,|\,t_{\parallel}\left(\frac{1}{2},\frac{1}{2}\right)\right]
\right\}$. The $\mathcal{M}_{\varphi}$,  $\mathcal{M}_{\varphi^{\prime}}$ symmetries, altermagnetic spin-splitting, and spin degeneracy at  $\Gamma$-point remain preserved, as evidenced by band structure shown in Fig.\ref{bilayer}(f,j). Within the range between, $t_{\parallel}(0,0)$ and $t_{\parallel}\left(\frac{1}{2},\frac{1}{2}\right)$, Mn$_2$WS$_4$ configuration belongs to the $Cm$ space group $C^{ \bar 1}m^{\infty_{ 1-10}m}1$), with symmetry operations classified into the sets $H \in
\{E,\; [\mathcal{P} \big|t_{\parallel}(t_a,t_b)]\}$ and ${G}-{H}
\in
\left\{
\mathcal{M}_{\varphi},\;
\left[\mathcal{C}_{2\varphi} \big| t_{\parallel}(t_a,t_b)\right]
\right\}$, where $t_a=t_b$. Consequently, altermagnetic spin-splitting originates from the breaking of  $[\mathcal{C}_2 \| \mathcal{P}]$ and $[\mathcal{C}_2 \| \mathcal{M}_{z}]$  symmetries, whereas the preserved $[\mathcal{C}_2 \| \mathcal{M}_{\phi}]$ symmetry gives rise to the degenerate valley at the $X$ and $X^{\prime}$ points carrying opposite spin characters.  The valence band maximum (VBM) at $X(X^{\prime})$ point is composed of spin-up (spin-down) band. The $\mathcal{M}_{\phi}$ symmetry operation transform $X(\pi,0)$ and $X^{\prime}(0,\pi)$ according to, $\widehat{\mathcal{M}_{\phi}}|X(X')\rangle = \widehat{\mathcal{M}_{\phi}}|X'(X)\rangle$. Let $|\psi_X\rangle$ represent the Bloch eigenstate of the Hamiltonian $\widehat{H}$ at $X$ point associated with the energy eigenvalue $E_{X}$ can be expressed as, $\widehat{H}|\psi_X\rangle = E_X |\psi_X\rangle$. Because $\widehat{H}$ commute with the $\widehat{\mathcal{M}_{\phi}}$ operator, $\widehat{H}\widehat{\mathcal{M}_{\phi}} =  \widehat{\mathcal{M}_{\phi}}\widehat{H}$. The $\widehat{\mathcal{M}_{\phi}}$ apply over eigenvalue equation as $\widehat{\mathcal{M}_{\varphi}}\widehat{H}|\psi_X\rangle = \widehat{H} \left(\widehat{\mathcal{M}_{\varphi}}|\psi_X\rangle\right) = \widehat{\mathcal{M}_{\varphi}} \left(E_X|\psi_X\rangle\right) = E_X \left(\widehat{\mathcal{M}_{\varphi}}|\psi_X\rangle\right)$. This clearly demonstrates that $\widehat{\mathcal{M}_{\varphi}}|\psi_X\rangle$ constitute another eigenstate possessing the same energy eigenvalue $E_X$. A similar treatment at the $X^{\prime}$ point gives $E_X = E_X^{\prime}$. Hence, the energy degeneracy at $X$ and $X^{\prime}$ points depicted in band structure of AB-bilayer is naturally explained by the underlying symmetry. 

In the second category $t_{\parallel}(t_a,t_b)$, characterized by $t_a \neq t_b$, the AC$_1$- and AC$_2$-bilayers are derived from the AA-bilayer through relative layer displacement along the fractional directions
$t_{\parallel}\left(\frac{1}{2},0\right)$ and
$t_{\parallel}\left(0,\frac{1}{2}\right)$, respectively (see Fig.\ref{bilayer}(c,d)). As the system evolved from $\left(0,0 \right)$ to $\left(0,\frac{1}{2}\right)$ and $\left(\frac{1}{2},0\right)$, Mn$_2$WS$_4$ undergoes a phase transition from the $P\bar{4}2m$ (spin space group: $P^{\bar 1}\bar{4}^{1}2^{\bar 1}m^{\infty 100m}1$) space group to $P222_1m$ (spin space group: $P^{ 1}{2}^{1}{2}^{1}2_{1}^{\infty 100m}1$) space group. The associated symmetry operations are categorized into the sets $H \in \left\{
E,\,[\mathcal{P}\big|\tau],\,[\mathcal{M}_z\big|\tau],\,\mathcal{C}_{2z}\right\}$, where $\tau=\frac{1}{2}a$ or $\tau=\frac{1}{2}b$ and ${G}-{H}=\varnothing$. For intermediate configuration between, $t_{\parallel}(0,0)$ and $t_{\parallel}\left(\frac{1}{2},0\right)$ or
$t_{\parallel}\left(0,\frac{1}{2}\right)$, the system is described by the
$P^{1}2^{\infty_{100}m}1$ spin space group, with symmetry operations classified into the sets $H \in
\{E,\; [\mathcal{P}\big| t_{\parallel}(t_a,t_b)]\}$, where $t_a \neq t_b$ and ${G}-{H}=\varnothing$. 

\begin{table*}[t!]
	\centering
	\tiny % Font size reduced to fit dense parameters on one vertical page
	\renewcommand{\arraystretch}{1.6} % Keeps math elements legible
	\setlength{\tabcolsep}{3pt} % Reduces padding between columns to maximize space
	
	\caption{Symmetry classification of different Mn$_2$WS$_4$ bilayer stacking configurations.}
	\label{tab:stacking}
	
	% Custom column definitions for optimal horizontal distribution
	\newcolumntype{C}[1]{>{\centering\arraybackslash}p{#1}}
	\newcolumntype{Y}{>{\centering\arraybackslash}X}
	\newcolumntype{L}{>{\raggedright\arraybackslash}X}
	
	% Total width is locked strictly to the text width of the vertical page
	\begin{tabularx}{\textwidth}{|C{1.8cm}|C{1.2cm}|C{1.8cm}|C{0.9cm}|C{0.8cm}|Y|L|}
		\hline
		
		\textbf{Mn$_2$WS$_4$ Sliding}
		&
		\textbf{Space Group}
		&
		\textbf{Spin Space Group}
		&
		\textbf{Phase}
		&
		\textbf{\makecell[c]{NRSS\\at $\Gamma$}}
		&
		\textbf{Spin-Laue Group}
		&
		\textbf{Comments}
		\\
		\hline
		
		AA-configuration
		&
		$P\bar{4}2m$ \newline (111)
		&
		$P^{\bar 1}\bar{4}^{1}2^{\bar 1}m^{\infty 100m}1$ \newline (111.16.1.1.L)
		&
		AM
		&
		\ding{55}
		&
		$\begin{aligned}
		H &\in \{E,\mathcal{C}_{2z},\mathcal{P},\mathcal{M}_{z}\},\\
		G-H \in & \{\mathcal{M}_{\varphi},\mathcal{M}_{\varphi'},
		\mathcal{C}_{2\varphi},\mathcal{C}_{2\varphi'}\}
		\end{aligned}$
		&
		Absence of $\mathcal{PT}$ symmetry; $[\mathcal{C}_2 \| \mathcal{C}_{4z}]$ spin symmetries serve as connectors that enforce altermagnetic spin splitting.
		\\
		\hline
		
		AB-configuration
		&
		$P\bar{4}2_1m$ \newline (113)
		&
		$P^{ 1}\bar{4}^{1}2^{1}_{1}m^{\infty 100m}1$ \newline (113.18.1.1.L)
		&
		AM
		&
		\ding{55}
		&
		$
		\begin{aligned}
		H \in \Bigl\{
		&E,\mathcal{C}_{2z},\\
		&\left[\mathcal{P}\Big|t_{\parallel}\!\left(\frac12,\frac12\right)\right],\\
		&\left[\mathcal{M}_{z}\Big|t_{\parallel}\!\left(\frac12,\frac12\right)\right]
		\Bigr\},\\[4pt]
		G-H \in \Bigl\{
		&\mathcal{M}_{\varphi},\mathcal{M}_{\varphi'},\\
		&\left[\mathcal{C}_{2\varphi}\Big|t_{\parallel}\!\left(\frac12,\frac12\right)\right],\\
		&\left[\mathcal{C}_{2\varphi'}\Big|t_{\parallel}\!\left(\frac12,\frac12\right)\right]
		\Bigr\}
		\end{aligned}
		$
		&
		Absence of $\mathcal{PT}$ symmetry; $[\mathcal{C}_2 \| \mathcal{M}_{\phi}]$ spin symmetries serve as connectors that enforce altermagnetic spin splitting.
		\\
		\hline
		
		AC$_1$-configuration
		&
		$P222_1$ \newline (17)
		&
		$P^{ 1}{2}^{1}{2}^{1}2_{1}^{\infty 100m}1$ \newline (17.17.1.1.L)
		&
		Quasi-AM
		&
		\ding{51}
		&
		$\begin{aligned}
		H \in & \{E,C_{2z},[\mathcal{P}|\tau],[\mathcal{M}_{z}|\tau]\},\\
		G-H \in & \varnothing
		\end{aligned}$
		&
		Absence of $\mathcal{PT}$ symmetry; spin symmetry operation no longer survives, removing protection against splitting at $\Gamma$ leads to SST-IV spin splitting, coupled to $(0,\frac12)$ through $[\mathcal{C}_2 \| \mathcal{M}_{1\bar 10}]$ symmetry operation.
		\\
		\hline
		
		AC$_2$-configuration
		&
		$P222_1$ \newline (17)
		&
		$P^{ 1}{2}^{1}{2}^{1}2_{1}^{\infty 100m}1$ \newline (17.17.1.1.L)
		&
		Quasi-AM
		&
		\ding{51}
		&
		$\begin{aligned}
		H \in & \{E,C_{2z},[\mathcal{P}|\tau],[\mathcal{M}_{z}|\tau]\},\\
		G-H \in & \varnothing
		\end{aligned}$
		&
		Absence of $\mathcal{PT}$ symmetry; spin symmetry operation no longer survives, removing protection against splitting at $\Gamma$, leads to SST-IV spin splitting, coupled to $(\frac12,0)$ through $[\mathcal{C}_2 \| \mathcal{M}_{1\bar 10}]$ symmetry operation.
		\\
		\hline
		
	\end{tabularx}
\end{table*}

Compared with $t_a = t_b$ configuration, the $G-H$ set is empty for $t_a \neq t_b$, indicating that the  $t_{\parallel}\left(\frac{1}{2},0\right)$ and  $t_{\parallel}\left(0,\frac{1}{2}\right)$ sliding operations do not fully satisfy the AM condition. The lack of $[\mathcal{C}_2 \| \mathcal{M}_{\phi}]$ symmetry is evident from the band structures in Fig.\ref{bilayer} (g) and (h), where spin-split bands along $\Gamma \rightarrow X \rightarrow M$ path differ from those along  $\Gamma \rightarrow X'\rightarrow M'$. Previous studies have described such behaviour as VP, however, here it originates from coupled quasi altermagnetic states. The AC$_1$ and AC$_2$ spin-sublattice configurations are related through the $[\mathcal{C}_2 \| \mathcal{M}_{1\bar 10}]$ symmetry operation in real space, which also establishes their correspondence in $k$-space. Accordingly, spin splitting along $\Gamma \rightarrow X \rightarrow M$ in AC$_1$ state maps onto the $\Gamma \rightarrow X'\rightarrow M'$ in AC$_2$ state, and vice versa, through $[\mathcal{C}_2 \| \mathcal{M}_{1\bar 10}]$ symmetry operation. This demonstrates a reversible spin-valley character between AC$_1$ and AC$_2$ states and suggests a unified description of different spin-split configurations. Furthermore, these coupled quasi-altermagnetic states exhibiting antiferromagnetic-like configurations consist of opposite spin sublattices that are not related by either proper/improper rotational symmetry operations. Similar to altermagnets, these systems display non-relativistic spin splitting along non-$\Gamma$ momentum paths, however, they further exhibit spin-resolved band splitting at the $\Gamma$-point (Fig.\ref{bilayer}(k) and (l)). Importantly, these coupled quasi altermagnetic states can be manipulated by non-magnetic perturbations such as sliding, strain, and multiferroic ordering, provided that the  $[C_2 \| O_s]$ orthogonal symmetry is restricted to specific stacking arrangements. Hence, the observed coupled quasi altermagnetic states resemble paired ferroelectric altermagnetic states, where spin flipping is achieved without any intrinsic electric polarization. A complete classification of the space group, spin-space group, magnetic phase, non-relativistic spin splitting at the $\Gamma$ point, corresponding spin-Laue group, and relevant remarks for all four sliding configurations is provided in Tab. \ref{tab:stacking}. The Mn-d orbital-projected electronic band structures were calculated for all four configurations (see Fig. \ref{bilayer_orbital}(a-d)). Among the five Mn-d orbitals, the d$_{x^2-y^2}$, d$_{xz}$, and d$_{yz}$ states exhibit the most significant contributions and are therefore analyzed in detail. Filled and open circles denote $\uparrow$ and $\downarrow$ states, respectively. In all configurations, the valence bands along $\Gamma$-M ($\Gamma$-M$^{\prime}$) direction are predominantly derived from  d$_{x^2-y^2}$ $\downarrow$ orbital. Along X-M and X$^{\prime}$-M$^{\prime}$ paths, the upper two bands mainly originate from  d$_{yz}$ $\uparrow$ and d$_{xz}$ $\uparrow$ orbitals, respectively. Meanwhile, the lower two bands are primarily contributed by the  d$_{x^2-y^2}$ $\uparrow$ and  d$_{x^2-y^2}$ $\downarrow$ orbitals, respectively.

\begin{figure}[ht!]
	\centering
	\includegraphics[width = 8cm]{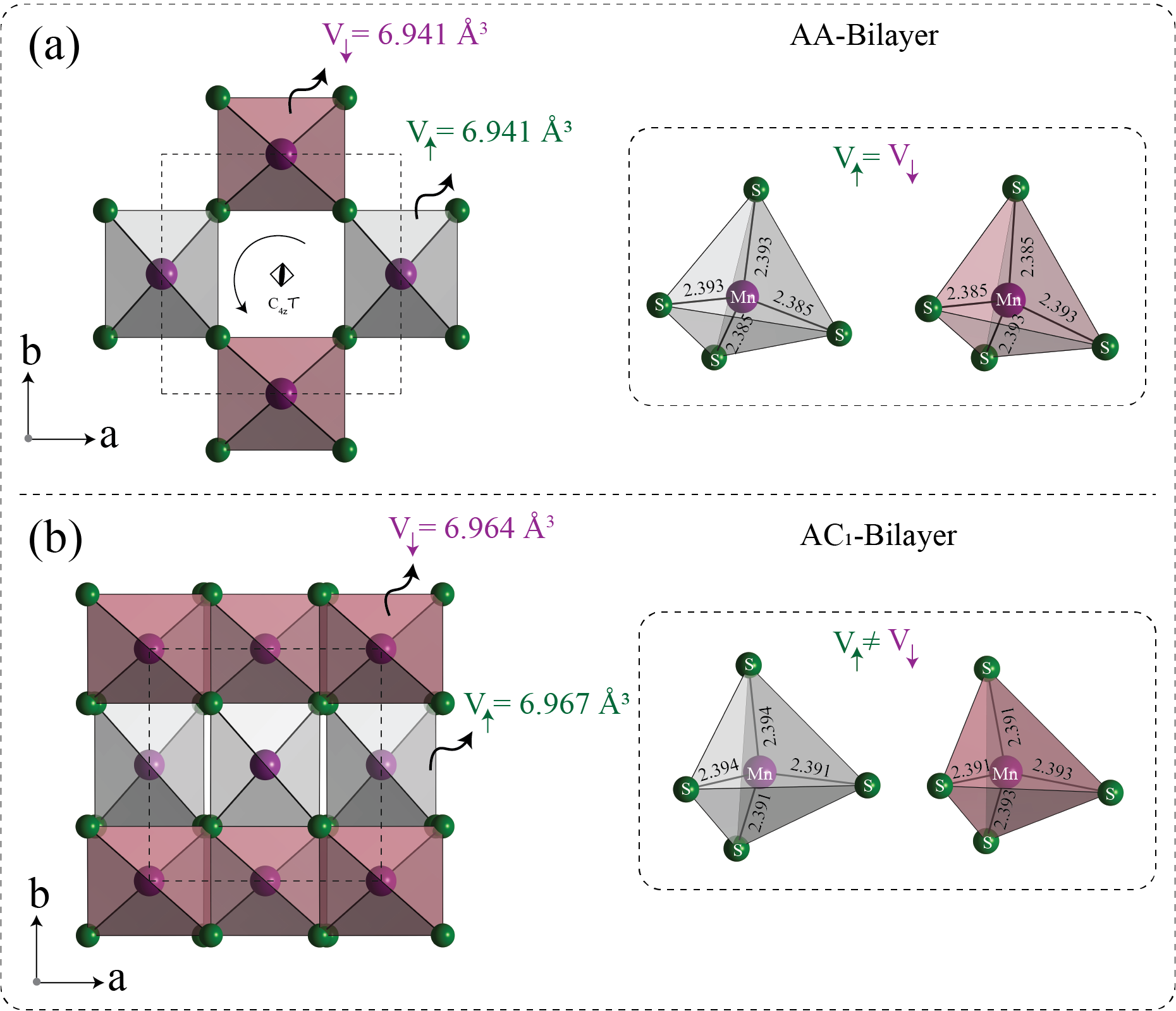}
	\caption{\label{tetra}Structural comparison of tetrahedra formed by Mn atoms in the $\uparrow$ and $\downarrow$ sublattices for the (a) AA and (b) AC$_1$ bilayers. Grey and light-pink colors denote the $\uparrow$ and $\downarrow$ tetrahedra, respectively. Separate views of the tetrahedra and the corresponding Mn-S bond lengths are highlighted beside the crystal structures.
	 }
\end{figure}

The sliding induced lifting of band degeneracy at $\Gamma$ point can be further understood from a structural perspective. Fig. \ref{tetra} (a) and (b) show relaxed crystal structures of AA- and AC$_1$-stacked bilayers, respectively. The grey and light-pink tetrahedra are formed by Mn atoms belonging to $\uparrow$ and $\downarrow$ sublattices, respectively. We calculated volumes of tetrahedra formed by $\uparrow$ and $\downarrow$ Mn atoms to quantify the structural asymmetry between two magnetic sublattices. In the AA bilayer, both tetrahedra possess identical volumes (V$_{\uparrow}$=V$_{\downarrow}$ = 6.941 $\mathrm{\AA}^3$). Indicating that the  same local environments in two spin channels (consistent with the definition of altermagnetism). As a result, $\Gamma$ point remains invariant under proper rotation, preserving the band degeneracy at $\Gamma$. In contrast, AC$_1$ bilayer exhibits a finite difference between V$_{\uparrow}$ and V$_{\downarrow}$ tetrahedra having $\Delta$ V = V$_{\uparrow}$ - V$_{\downarrow}$ $\approx$ 3 $\times 10^{-3}\mathrm{\AA}^3$. Although this volume difference is relatively small, it is sufficient to break equivalence between two magnetic sublattices and lift degeneracy at $\Gamma$ point (consistent with the definition of quasi-altermagnetism). The volume disproportionation also generates a local structural void or distortion within the crystal lattice. The volume disproportionation is further reflected in Mn-S bond lengths. A similar analysis reveals that the AB bilayer also possesses equal tetrahedral volumes, whereas the AC$_2$ configuration exhibits unequal tetrahedral volumes. Such a mechanism may also operate in doped altermagnets, where dopant-induced variations in bond lengths and local coordination environments can create inequivalent $\uparrow$ and $\downarrow$ sublattices \cite{yuan2024nonrelativistic,devaraj2026unlocking}.

\begin{figure}[ht!]
	\centering
	\includegraphics[width = 8cm]{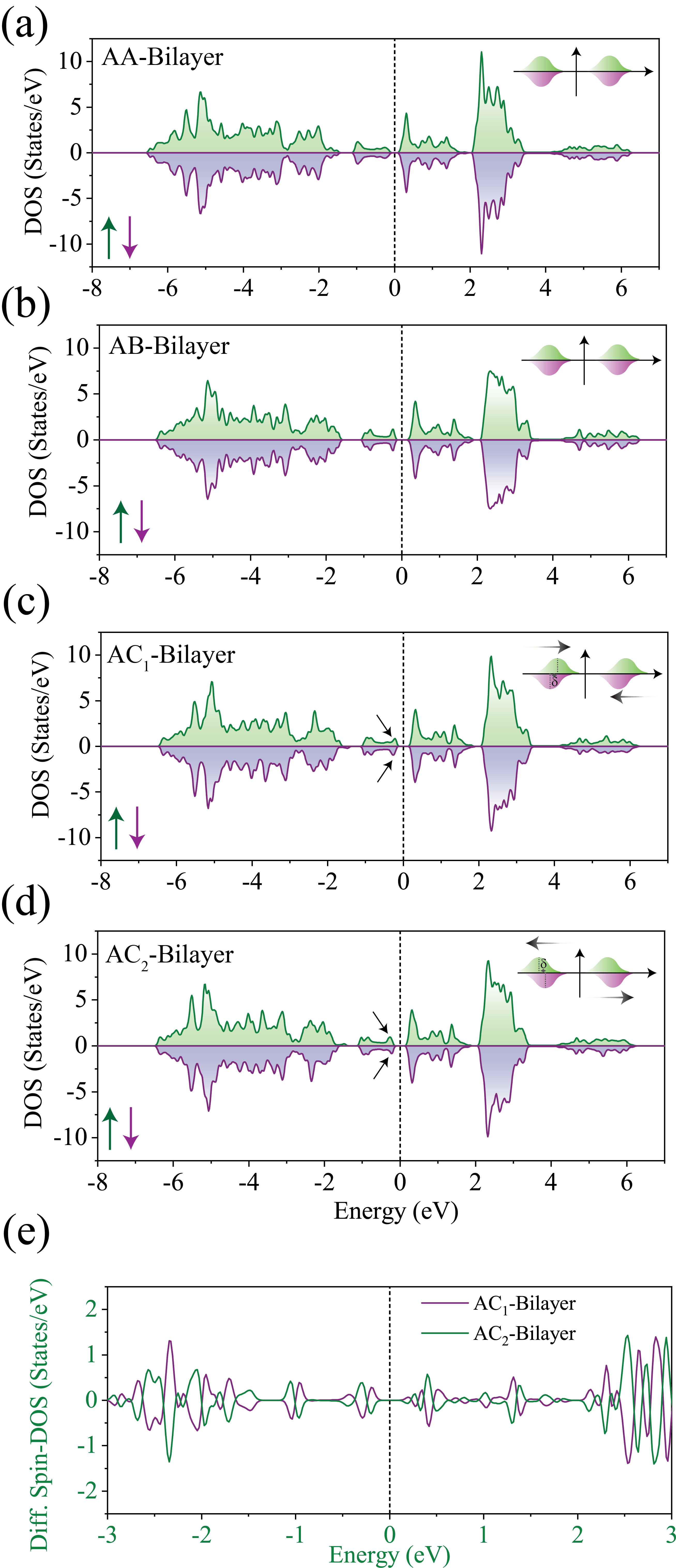}
	\caption{\label{pdos} Spin-resolved DOS for the (a) AA, (b) AB, (c) AC$_1$, and (d) AC$_2$ bilayer configurations. Insets schematically illustrate the relative positions of the spin-dependent band edges in each configuration. In the AA and AB bilayers, the $\uparrow$ and $\downarrow$ DOS exhibit identical band-edge positions, consistent with the exactly same $\uparrow$ and $\downarrow$ sublattices. In contrast, the AC$_1$ and AC$_2$ configurations display opposite shifts of the spin-dependent band edges, indicating sliding-integrated signatures of spin-split electronic states. The direction of the corresponding band-edge shifts are highlighted by arrows in the DOS plots.(e) Difference between the spin-resolved DOS of the AC$_1$ and AC$_2$ configurations. The nearly perfect sign reversal demonstrates the coupled configurations.}
\end{figure}

The behaviour of different stacking configurations was analyzed in terms of the net magnetization, defined as $M=\mu_B\int_{-\infty}^{E_F} \left[D^{\uparrow}(E)-D^{\downarrow}(E)\right]\,dE$, where  $E_F$ is Fermi, $\mu_B$ is Bohr magneton, and $D^{\uparrow}(E)$ and $D^{\downarrow}(E)$ corresponds to the $\uparrow$ and $\downarrow$ DOS, respectively. The AA and AB bilayer structures exhibit symmetric integrated DOS ($iDOS$) for both spin channels (see Fig.\ref{pdos}(a,b)). As a result, the band-edge positions remain degenerate for spin-selective excitations, giving rise to antiferromagnet-like demagnetization behaviour. For AC$_1$ and AC$_2$ configurations, however,${G}-{H}=\varnothing$, symmetry relating the spin sublattices is broken, although the condition $iDOS^{\uparrow}(E_F)=iDOS^{\downarrow}(E_F)$ still ensures vanishing net magnetization. Such compensation is restricted to semiconducting or insulating phases, metallic systems with $iDOS^{\uparrow}(E_F) \neq iDOS^{\downarrow}(E_F)$ develop a finite magnetic moment. Moreover, opposite interlayer sliding in AC$_1$ and AC$_2$ produces an asymmetric spin-resolved DOS distribution (Fig.\ref{pdos}(c,d)), as shown schematically in the inset. The distinct band-edge positions for the two spin channels yield unequal spin-dependent band gaps, which drive an uncompensated spin-transfer process and ultimately result in asymmetric demagnetization. Fig.\ref{pdos}(e) shows the difference in spin DOS, $iDOS^{\uparrow}(E_F)-iDOS^{\downarrow}(E_F)$, for the AC$_1$ and AC$_2$ bilayers, revealing opposite difference spin-polarization characteristics in the two configurations. In contrast, the AA- and AB-bilayers exhibit zero difference in spin DOS owing to complete spin compensation. To obtain further insight into the coupled quasi-altermagnetic phases, we examined spin-resolved Fermi surfaces corresponding to AA-, AC$_1$-, and AC$_2$-configurations.

\begin{figure}[ht!]
	\centering
	\includegraphics[width = 14cm]{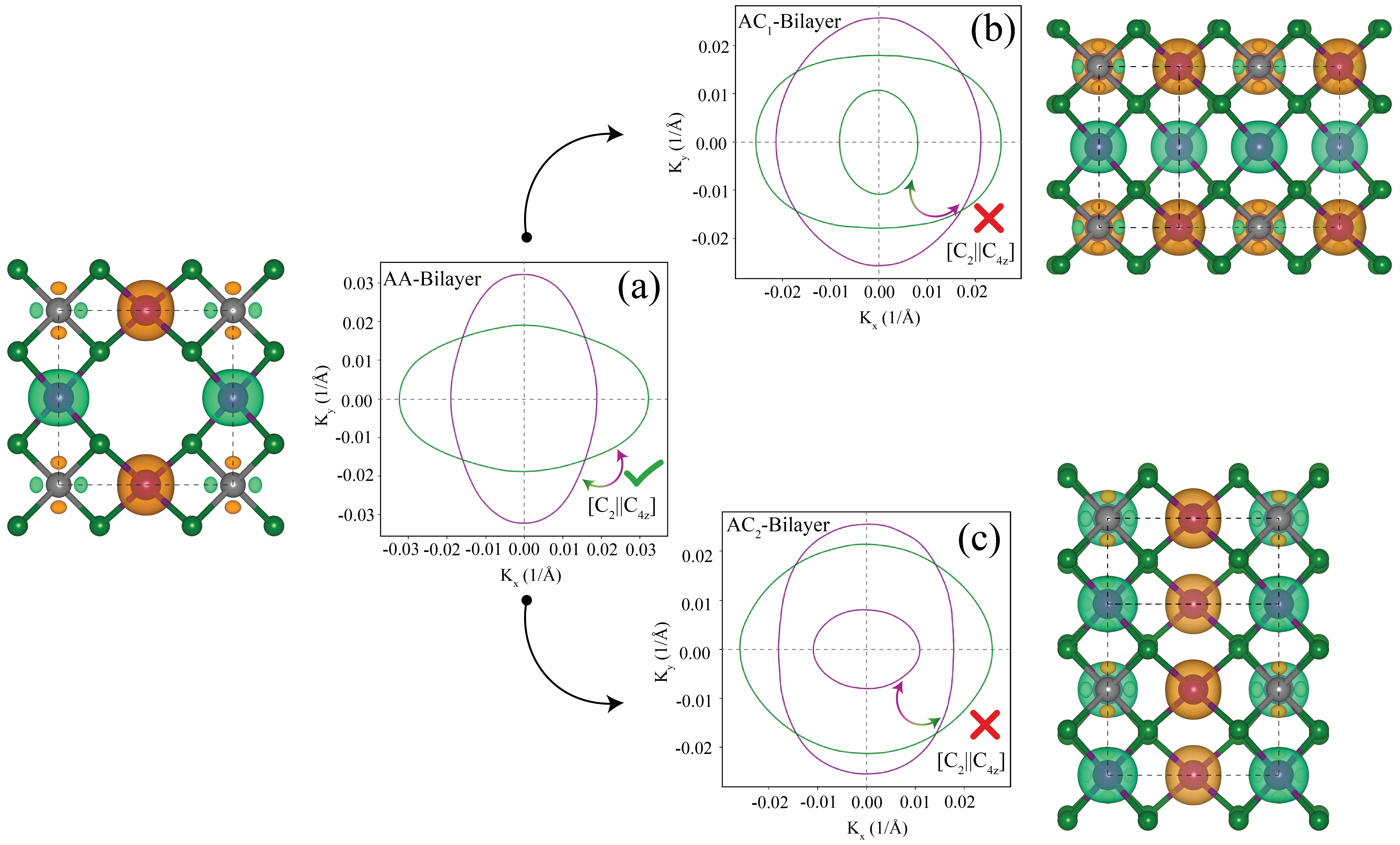}
	\caption{\label{fermi}Spin-resolved Fermi surfaces of the (a) AA, (b) AC$_1$, and (c) AC$_2$ bilayer configurations. The corresponding spin-resolved charge densities are shown alongside each Fermi surface. In the AA bilayer, the $\uparrow$ and $\downarrow$ Fermi surfaces are related by the combined symmetry operation $[\mathcal{C}_2 \| \mathcal{C}_{4z}]$, marked with a green check mark, reflecting the characteristic d-wave altermagnetic spin splitting and existence of a valid opposite-spin sublattice connector. In contrast, the AC$_1$ and AC$_2$ configurations lack a valid opposite-spin sublattice connector. The $\uparrow$ Fermi surface (green) of AC$_1$ configuration  is transformed into the $\downarrow$ Fermi surface (purple) of AC$_2$ configuration through the corresponding rotational symmetry operation.}
\end{figure}

For the AA bilayer (see Fig.\ref{fermi} (a)), the Fermi surfaces along the $K_x-K_y$ planes are symmetric in shape but carry opposite spin polarization, leading to the $[\mathcal{C}_2 \| \mathcal{C}_{4z}]$ connect opposite spin sublattices. Unlike the AA configuration, the AC$_1$ and AC$_2$ coupled quasi-altermagnetic states exhibit intrinsically \textit{asymmetric} spin-up and spin-down Fermi surfaces that cannot be transformed into one another by any symmetry operation, consistent with their electronic band structures. However, the Fermi surfaces of AC$_1$ and AC$_2$ remain related through the  $[\mathcal{C}_2 \| \mathcal{M}_{1\bar 10}]$  symmetry operation (Fig.\ref{fermi} (b,c)). In addition, charge density distributions around Mn atoms for all three configurations are presented alongside the corresponding Fermi surfaces. Besides the $\mathcal{M}_{\phi}$ symmetry changes induced by sliding (preserved in AA- and AB-stacked bilayer configurations and broken in AC$_1$ and AC$_2$-configurations), the orientation of easy magnetization axis can further modify the $\mathcal{M}_{\phi}$ symmetry. Zhang et al. \cite{zhang2025strain} reported that the reorientation of the easy axis, governed by SOC, significantly affects the $\mathcal{M}_{\phi}$ symmetry properties. The $\mathcal{M}_{\phi}$ symmetry is preserved for in-plane spin configurations when (i) the magnetic moment of atom 1(atom 2) are aligned along the $[001]([00\bar1])$ directions and (ii) both magnetic moments remain parallel to the mirror plane. 

\begin{figure}[ht!]
	\centering
	\includegraphics[width = 16cm]{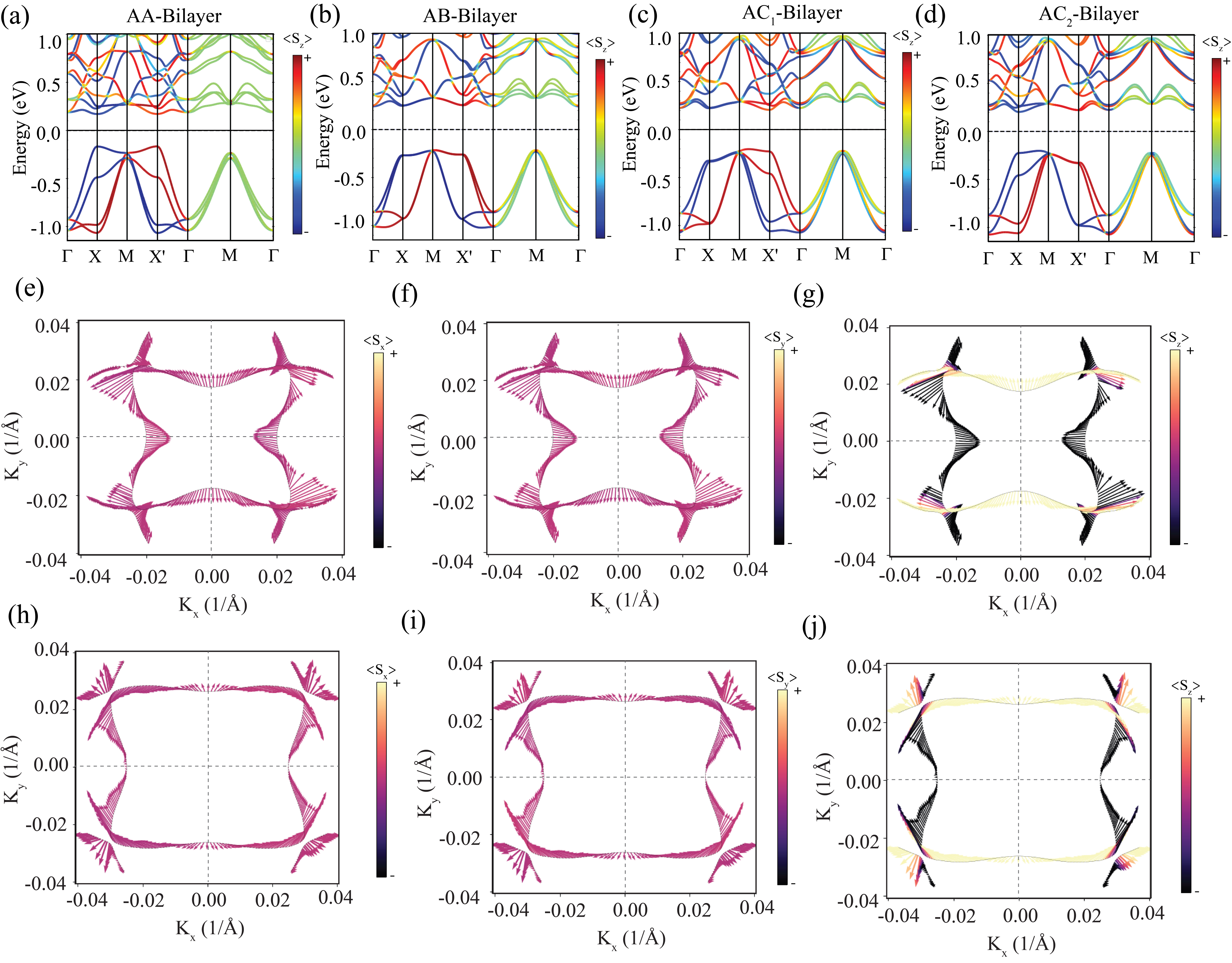}
	\caption{\label{soc}Spin-resolved electronic band structures projected onto the $S_z$ component in the presence of SOC for the (a) AA, (b) AB, (c) AC$_1$, and (d) AC$_2$ bilayer configurations. Color scale represents magnitude and sign of spin polarization associated with $S_z$ component.
		(e-g) and (h-j) Spin textures projected onto the $S_x$, $S_y$, and $S_z$ components for the AC$_1$ and AC$_2$ bilayer configurations, respectively. }
\end{figure}

\begin{figure}[ht!]
	\centering
	\includegraphics[width = 16cm]{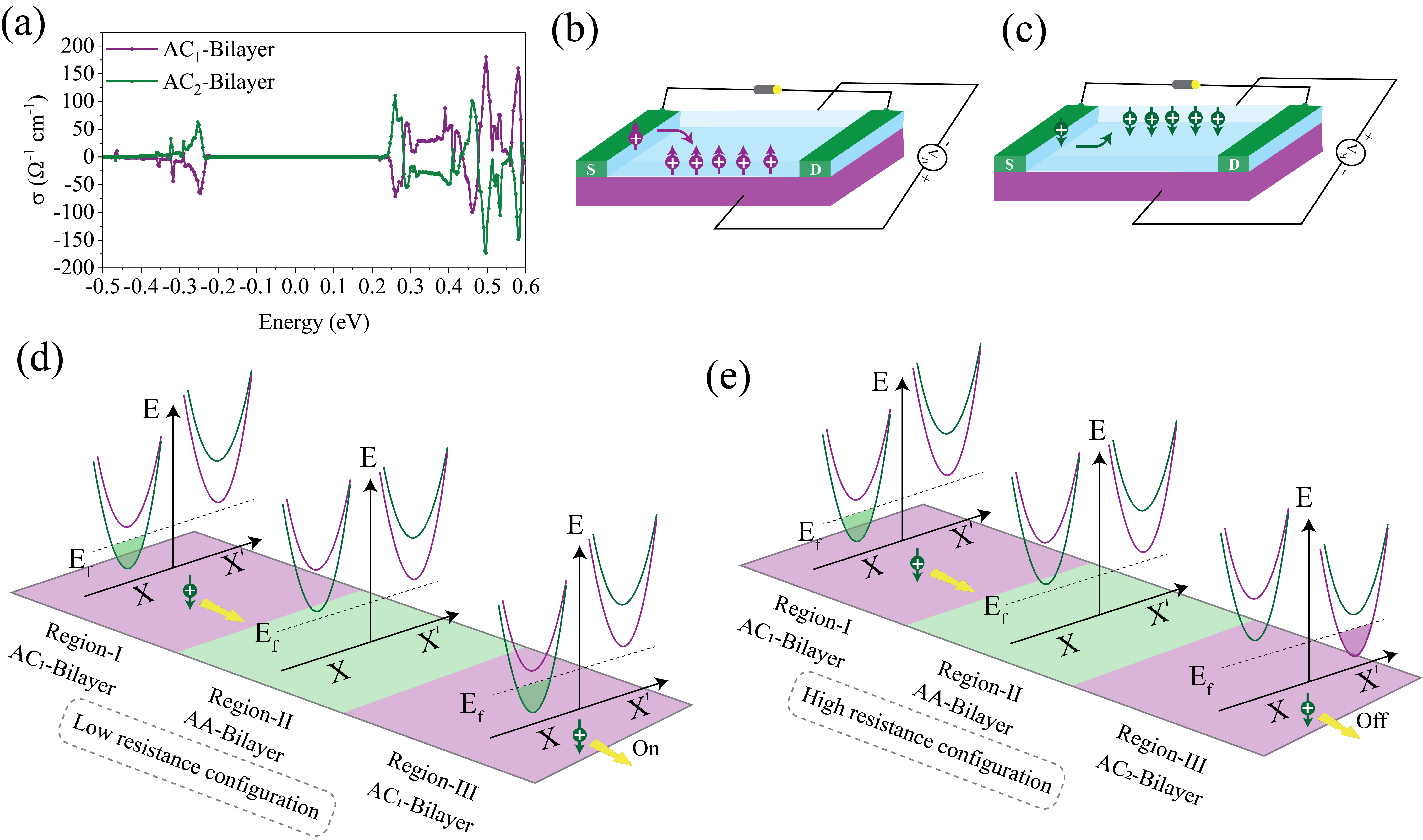}
	\caption{\label{AHC}(a) AHC of the AC$_1$ and AC$_2$ bilayer configurations, shown in purple and green, respectively. (b,c) Schematic illustrations of anomalous VHE in AC$_1$ and AC$_2$ bilayer under hole doping and in the presence of an in-plane electric field. (d,e) Proposed tunneling magnetoresistance(TMR)-like device architectures based on the coupled quasi-altermagnetic bilayers. (d) Low-resistance (parallel) configuration, consisting of a hole-doped AC$_1$ (or AC$_2$) $\oplus$ AA $\oplus$ hole-doped AC$_1$ (or AC$_2$) electrode. (e) High-resistance (antiparallel) configuration, composed of a hole-doped AC$_1$ $\oplus$ AA $\oplus$ hole-doped AC$_2$ electrode. }
\end{figure}

Conversely, mirror symmetry is broken when spins are oriented perpendicular to mirror plane or contain a finite out-of-plane component. Effect of SOC on all four configurations is shown in Fig.\ref{soc} (a-d), where SOC induced lifting of band degeneracy is clearly observed. In the absence of SOC, the AA- and AB-bilayer configurations exhibit spin-degenerate bands along the $\Gamma \rightarrow M$  path by virtue of the  $[\mathcal{C}_2 \| \mathcal{M}_{\phi}]$ spin symmetry. Upon inclusion of SOC, this degeneracy is lifted, resulting in spin-split bands. In contrast, the  $[\mathcal{C}_2 \| \mathcal{M}_{\phi}]$ symmetry is already broken in the AC$_1$ and AC$_2$ configurations, leading to lifted band degeneracy along the $\Gamma \rightarrow M$ direction even without SOC. The inclusion of SOC further enhances the spin-splitting effect in these configurations. Fig.\ref{soc} (e-g) and (h-j) illustrate the spin textures of the AC$_1$ and AC$_2$ bilayers under the influence of SOC, respectively. For both configurations, the in-plane spin components S$_x$ and S$_y$ remain nearly negligible across most of the electronic bands, indicating dominance of out-of-plane spin component S$_z$ in spin texture. The near-zero S$_x$ and S$_y$ components are represented by the pink color in the color scale. In contrast, the S$_z$ component exhibits significant spin polarization. Furthermore, the S$_z$ component along $\Gamma$-X and $\Gamma$-X$^{\prime}$ displays opposite orientations, corresponding to $\frac{\hbar}{2}$ and $-\frac{\hbar}{2}$, respectively, as highlighted by the black and light-yellow colors. With the inclusion of SOC, the AA- and AB-stacked bilayers exhibit $\mathcal{M}_{\phi}$ symmetry-protected altermagnetic spin-split bands, resulting in Berry curvature at $X$ and $X^{\prime}$ valley exhibit opposite sign but identical magnitude. This Berry curvature driven compensation gives rise to a VHE under an applied electric field, analogous to that observed in the monolayer system. In contrast, the AC$_1$ and AC$_2$ bilayer configurations display asymmetric Berry-curvature distributions in terms of magnitude as well as direction at the $X$ and $X^{\prime}$ valleys owing to the empty $G-H$ symmetry set, thereby producing an anomalous VHE (see Fig. \ref{AHC}(a)). Furthermore, the anomalous Hall conductivity in the AA- and AB-bilayer configurations becomes zero near the valence- and conduction-band edges, yielding pure valley and spin Hall currents without an accompanying charge Hall current \cite{sheoran2025spontaneous}. The application of uniaxial strain removes Berry-curvature equivalence between two valleys, stabilizing a valley-polarized state and inducing an anomalous VHE in the strained phase. In the anomalous VHE, an in-plane electric field generates a valley-dependent Hall current, causing transverse separation of charge carriers. Under a longitudinal electric field, the carriers acquire a transverse anomalous velocity given by $\mathbf{v}_{\perp}= \frac{e}{\hbar}\,[\mathbf{E}\times\mathbf{\Omega}(\mathbf{k})]$. Hole doping shifts Fermi level into valence band, causing valence-band maxima at the $X$ or $X^{\prime}$ valleys to intersect Fermi level. In the AC$_1$ state, an applied electric field drives spin-up holes toward lower edge of the sample, leading to carrier accumulation at the boundary and generating a positive transverse Hall voltage (schematically presented in Fig. \ref{AHC}(b)). In contrast, in the AC$_2$ state, spin-down holes migrate toward the upper edge under the same electric field, producing carrier accumulation at the opposite boundary and consequently a negative transverse Hall voltage (schematically presented in Fig. \ref{AHC}(c)).

These coupled quasi-altermagnetic configurations also provides a promising platform for designing sliding-controlled tunnelling magnetoresistance (TMR)-like devices in bilayer Mn$_2$WS$_4$. Unlike traditional TMR architectures that rely on external magnetic fields, the proposed mechanism exploits interlayer sliding to switch between quasi-altermagnetic states. Such sliding-induced transitions lead to complete reversal of spin polarization together with inversion of valley-selective transport behaviour. In the AC$_1$ configuration, $\uparrow$ electrons preferentially occupy the $X$ valley, whereas in the AC$_2$ configuration $\downarrow$ electrons are localized at $X^{\prime}$ valley. Therefore, AC$_1$ and AC$_2$ can act as bistable electrode states corresponding to low- and high-resistance transport configurations. Each device configuration consists of three distinct regions. The low-resistance (parallel) configuration (schematically illustrated in Fig. \ref{AHC}(d)) is composed of hole-doped AC$_1$ or AC$_2$ electrodes separated by an AA-stacked barrier region, i.e., hole-doped AC$_1$ (or AC$_2$) $\oplus$ AA $\oplus$ hole-doped AC$_1$ (or AC$_2$). The AA-stacked region functions as the tunnelling barrier, while the identical electrode configurations on both sides enhance electron tunnelling from left to right. Because the spin and valley degrees of freedom remain conserved during transport, this configuration exhibits a high transmission probability and therefore corresponds to the low-resistance state. By contrast, the high-resistance (antiparallel) configuration (schematically illustrated in Fig. \ref{AHC}(e)) consists of hole-doped AC$_1$ $\oplus$ AA $\oplus$ hole-doped AC$_2$. In this arrangement, the opposite electrode configurations and the inequivalent $X$ and $X^{\prime}$ valleys induce simultaneous spin and valley mismatch for electrons traversing the junction. This mismatch suppresses the tunnelling probability, thereby giving rise to the high-resistance state.

\begin{figure}[ht!]
	\centering
	\includegraphics[width = 16cm]{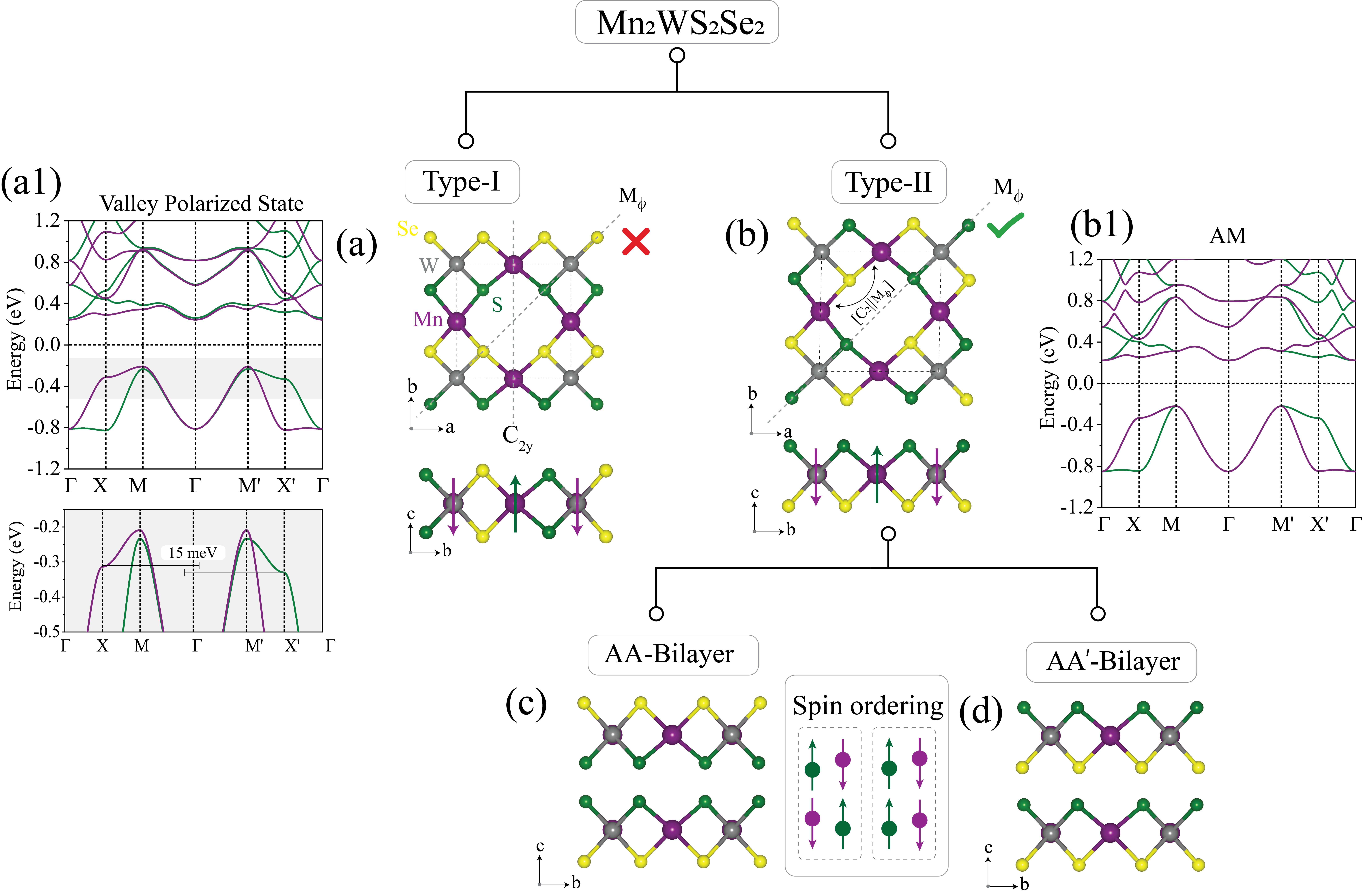}
	\caption{\label{janus}Diagram summarizing the structural evolution and magnetic phases of Mn$_2$WS$_2$Se$_2$. The top and side views of the crystal structures for (a) Type-I Mn$_2$WS$_2$Se$_2$ and (b) Type-II Mn$_2$WS$_2$Se$_2$ are shown. The preservation and breaking of the $\mathcal{M}_{\phi}$ symmetry are indicated by green check marks and red crosses, respectively. The existence of a valid opposite-spin sublattice connector in Type-II Mn$_2$WS$_2$Se$_2$ is explicitly highlighted. (a1,b1) Electronic band structures corresponding to Type-I and Type-II Mn$_2$WS$_2$Se$_2$, respectively. Type-I Mn$_2$WS$_2$Se$_2$ exhibits a valley-polarized electronic state, whereas Type-II Mn$_2$WS$_2$Se$_2$ realizes an altermagnetic phase. The Type-II Mn$_2$WS$_2$Se$_2$ structure is further classified into (c) AA and (d) AA$^{\prime}$ bilayer configurations. For each stacking geometry, two magnetic arrangements are considered, namely parallel $\uparrow$$\downarrow$/$\uparrow$$\downarrow$ and antiparallel $\uparrow$$\downarrow$/$\downarrow$$\uparrow$ spin alignments, as illustrated in the central panel. }
\end{figure}
 
Motivated by the symmetry analysis of Mn$_2$WS$_4$, we further construct Mn$_2$WS$_2$Se$_2$ system by partially replacing S atoms with Se atoms. Depending on the relative arrangement of S and Se atoms around the Mn sites, two distinct structural phases emerge: a non-Janus phase (type-I) (Fig. \ref{janus}(a))  and a Janus phase (type-II)(Fig. \ref{janus}(b)). These two phases crystallize in tetragonal structures belonging to the $P2$ (spin space group: $P^{1}2^{\infty 100m}1$) and $Cmm2$ (spin space group: $C^{\bar 1}{m}^{\bar 1}m^{1}2^{\infty_{1-10m}}1$) space groups, respectively. In the type-I phase, Mn-W layer is \textit{asymmetrically} enclosed by S-Se layers, leading to breaking inversion  while preserving $\mathcal{C}_{2y}$ and $\mathcal{C}_{2z}$ rotational symmetry. The broken $\mathcal{M}_{\phi}$ symmetry removes the equivalence between $X$ and $X^{\prime}$ valleys and consequently produces valley-polarized state. This behavior is consistent with earlier reports of valley polarization in tetragonal altermagnetic systems, through breaking of $\mathcal{C}_{4z}\mathcal{T}$ symmetry induced by uniaxial strain \cite{qi2024spin}. The calculated band structure of type-I configuration exhibits an energy splitting of $\approx$ 15 meV between $X$ and $X^{\prime}$ valleys (see Fig. \ref{janus}(a1)). The asymmetric distribution of S and Se atoms around Mn sites lifts Berry-curvature degeneracy of the two valleys, thereby stabilizing the valley-polarized phase and giving rise to an anomalous VHE. However, the type-I configuration does not possess any symmetry operation connecting the opposite magnetic sublattices, empty $G-H$ symmetry set. Therefore, this phase is excluded from further investigation and subsequent analysis is focused on the altermagnetic type-II Mn$_2$WS$_2$Se$_2$ (see Fig. \ref{janus}(b1)). 

In contrast to the Mn$_2$WS$_4$ bilayer configuration, type-II Mn$_2$WS$_2$Se$_2$ possesses a lower-symmetry structure, enabling a broader exploration of coupled quasi-altermagnetic states via modifications of the local environment arising from different interlayer stacking arrangements. Based on the fundamental stacking geometry, two distinct stacking configurations are possible: $AA$ (see Fig. \ref{janus}(c)) and $AA^{\prime}$ stacking (see Fig. \ref{janus}(d)). Two spin configurations were considered in each arrangement: (i) parallel alignment $\uparrow$$\downarrow$/$\uparrow$$\downarrow$ and (ii) antiparallel alignment $\uparrow$$\downarrow$/$\downarrow$$\uparrow$ \cite{gonzalez2025engineering}. 

\begin{figure}[ht!]
	\centering
	\includegraphics[width = 16cm]{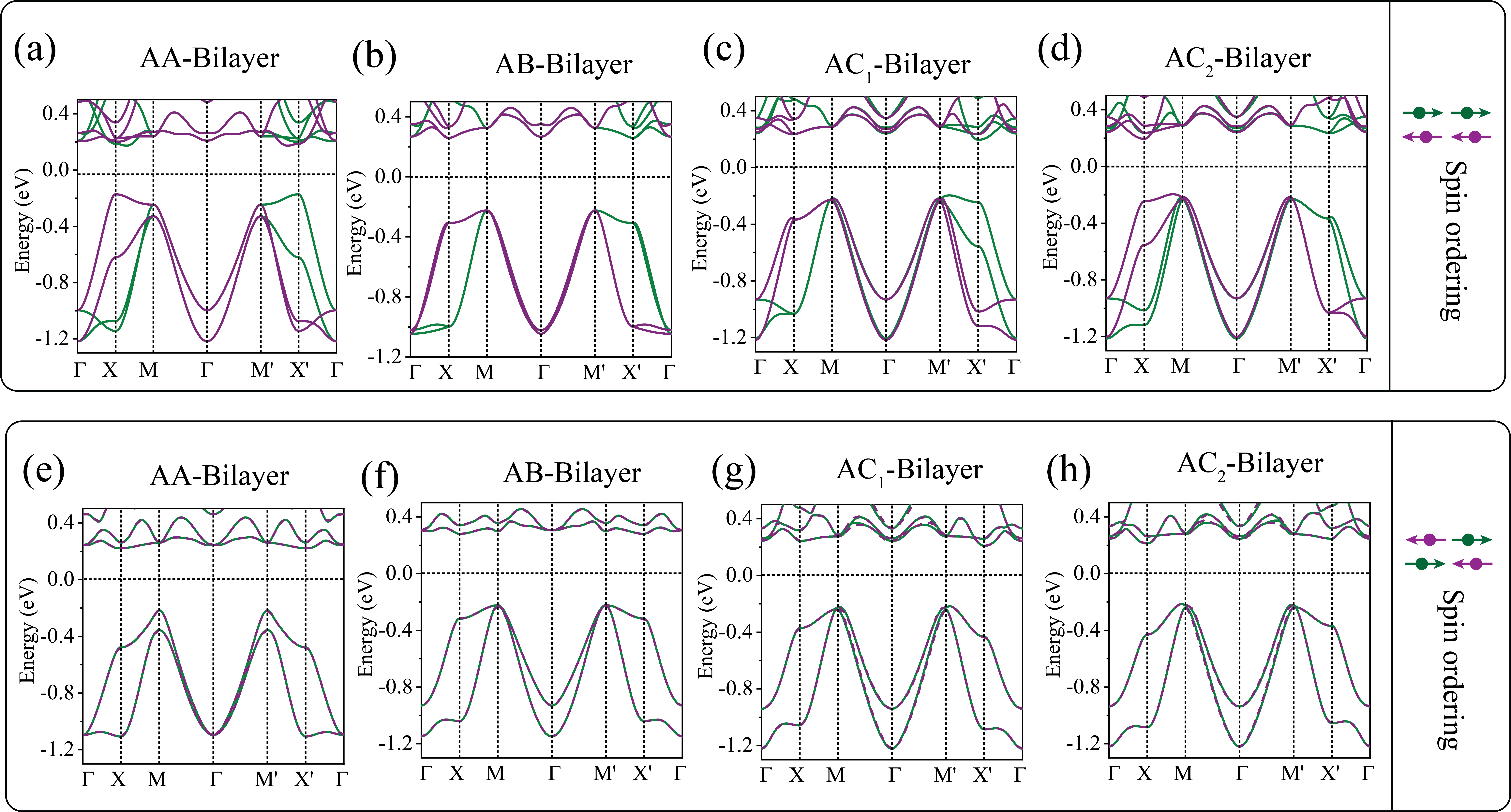}
	\caption{\label{AA}Spin-polarized electronic band structures of Type-II $AA$ Mn$_2$WS$_2$Se$_2$ bilayers for various stacking configurations and interlayer magnetic alignments. (a-d) AA, AB, AC$_1$, and AC$_2$ bilayers with parallel spin alignment, respectively. (e-h) Corresponding AA, AB, AC$_1$, and AC$_2$ bilayers with antiparallel spin alignment. }
\end{figure}

In AA-stacked Mn$_2$WS$_2$Se$_2$ structure, the upper monolayer is rotated by 180$^\circ$ relative to the lower monolayer before stacking, resulting in the preservation of both $\mathcal{M}_{z}$ and  $\mathcal{P}$ symmetry. In parallel spin alignment system exhibits AM behaviour due to the absence of $[\mathcal{C}_2 \| \mathcal{M}_{z}]$ and $\mathcal{PT}$ symmetry, while the opposite-spin sublattice connector remains intact (Fig. \ref{AA}(a)). Consequently, for the parallel spin configuration, Mn1-Mn4 and Mn2-Mn3 spin sublattices do not preserve the $\mathcal{M}_{z} \mathcal{U}$ and $\mathcal{PT}$ symmetries. Similar to AB-Mn$_2$WS$_4$, the AB-configuration for Mn$_2$WS$_2$Se$_2$ also preserves $\mathcal{M}_{\phi}$ symmetry during sliding, enforcing identical band dispersions for the $\uparrow$ and $\downarrow$ channels across the entire BZ, which is a hallmark of AM states (Fig. \ref{AA}(b)). Conversely, AC$_1$ and AC$_2$ configurations exhibit an empty $G-H$ set, resulting in coupled quasi-altermagnetic states (Fig. \ref{AA}(c,d)). In the antiparallel spin alignment, Mn1 and Mn3 (Mn2 and Mn4) spin sublattices are connected through the $\mathcal{M}_{z} \mathcal{U}$ and $\mathcal{PT}$ symmetry operations, producing fully compensated sublattices and consequently a degenerate antiferromagnetic band structure, see Fig. \ref{AA}(e). Under an external electric field, $\mathcal{PT}$ symmetry is broken, while $[\mathcal{C}_2 \| A]$ symmetry remains intact, resulting in the transformation of the antiferromagnetic state into an altermagnetic one \cite{qi2024spin,tian2025spin}. The resulting spin-splitting magnitude is proportional to the electric-field strength. Whereas spin-splitting reversal is controlled by the direction of electric field. Although the electric field breaks $\mathcal{PT}$ symmetry but it connects the $\psi_{+E}$ and $\psi_{-E}$ altermagnetic states arise by virtue of the positive and negative electric field, respectively. As a results, spin-splitting sign reverses with reversal of electric field direction, $\mathcal{PT}\psi_{+E}(s,\mathbf{k})=\psi_{-E}(-s,\mathbf{k})$. Recently, electric field induced AFM to AM transition has been demonstrated in several 2D compounds \cite{qi2024spin,tian2025spin}. Upon introducing sliding in the antiparallel arrangement, the AB configuration continues to preserve $\mathcal{M}_{\phi}$ symmetry, resulting in the retention of the AFM state (Fig. \ref{AA}(f)). In contrast, the AC$_1$ and AC$_2$ configurations display a small spin splitting along $\Gamma$-M path, as illustrated in Fig. \ref{AA}(g,h). 

\begin{figure}[ht!]
	\centering
	\includegraphics[width = 16cm]{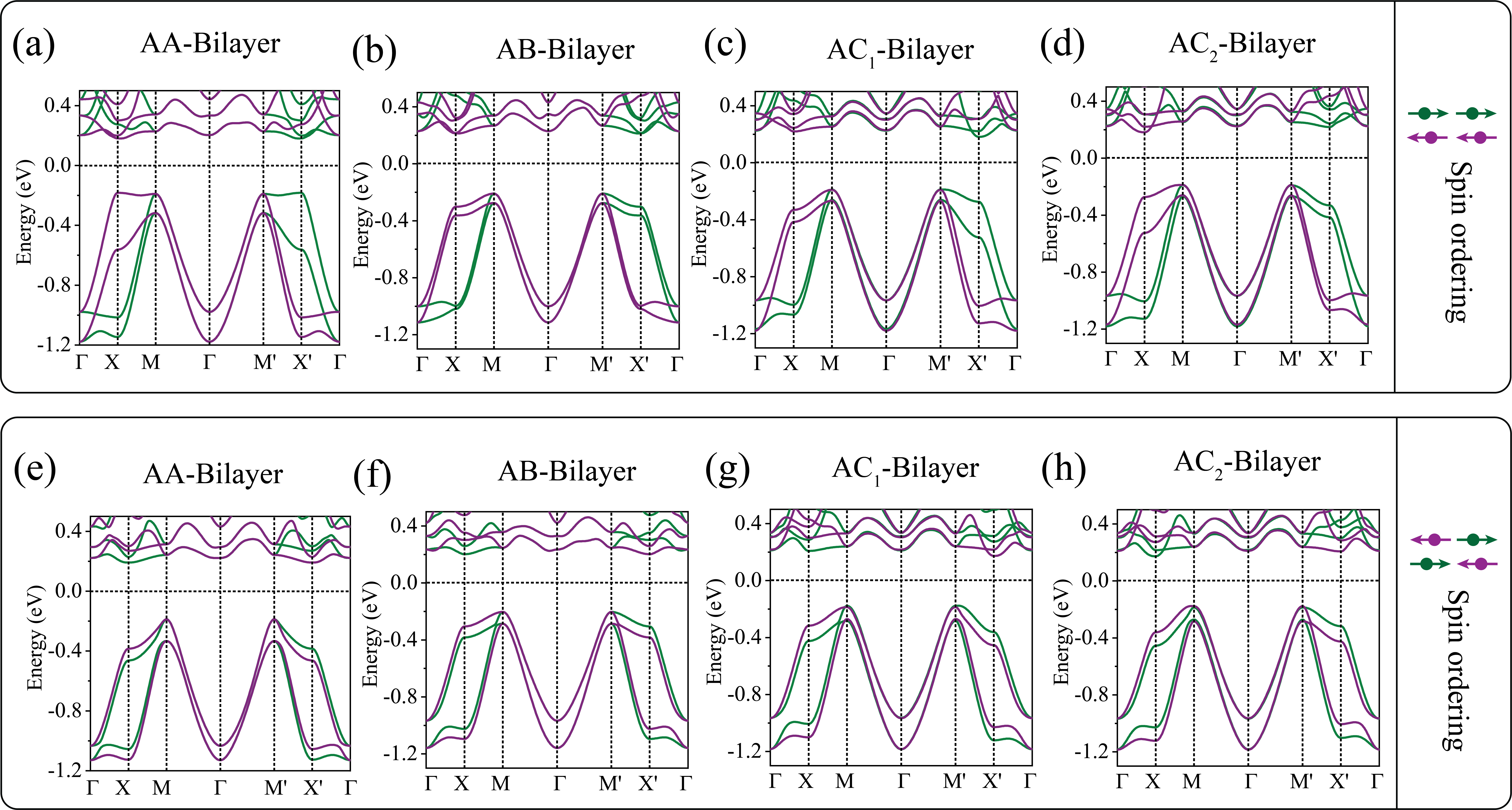}
	\caption{\label{AAP} Spin-polarized electronic band structures of Type-II $AA^{\prime}$ Mn$_2$WS$_2$Se$_2$ bilayers for various stacking configurations and interlayer magnetic alignments. (a-d) AA, AB, AC$_1$, and AC$_2$ bilayers with parallel spin alignment, respectively. (e-h) Corresponding AA, AB, AC$_1$, and AC$_2$ bilayers with antiparallel spin alignment.}
\end{figure}

In $AA^{\prime}$-stacked Mn$_2$WS$_2$Se$_2$ structure, the in-plane geometry remains identical in both layers, resulting in breaking of the $\mathcal{M}_{z}$ symmetry. The $AA^{\prime}$ stacking configuration modifies the symmetry landscape by eliminating both the $\mathcal{PT}$ and $\mathcal{M}_{z}$ symmetries. In parallel spin configuration, absence of $\mathcal{M}_{z} \mathcal{U}$ and $\mathcal{PT}$ symmetry operations relating the Mn1 and Mn2 (Mn3 and Mn4) spin sublattices gives rise to altermagnetic states in the AA and AB bilayers, as illustrated in Fig. \ref{AAP}(a,b). Similarly, quasi-altermagnetic states emerge in the AC$_1$ and AC$_2$ bilayers (see Fig. \ref{AAP}(c,d)), closely resembling the behavior of the Mn$_2$WS$_4$ monolayer. This originates from the simultaneous breaking of the relevant interlayer and intralayer symmetry operations. For antiparallel spin configuration, spin sublattices of Mn1 and Mn4 (Mn2 and Mn3)  are not related by $\mathcal{M}_{z} \mathcal{U}$ or $\mathcal{PT}$ symmetry operations. The same symmetry breaking also occurs between the intralayer Mn1 and Mn2 (Mn3 and Mn4) spin sublattices, resulting in an altermagnetic-like band structure, see Fig. \ref{AAP}(e). Furthermore, introduction of interlayer sliding preserves altermagnetic states in AB configuration, as well as the quasi-altermagnetic states in the AC$_1$ and AC$_2$ configurations, see Fig. \ref{AAP}(f-h).  

\section*{Conclusions}

In summary, we have established a comprehensive framework for understanding the sliding-induced interplay between altermagnetism and quasi-altermagnetic states by combining first-principles calculations, general stacking theory, and spin-Laue symmetry analysis. Our study identifies coupled quasi-altermagnetic states characterized by reversible type-IV nonrelativistic spin splitting, which can be effectively controlled through interlayer sliding. We establish a direct correspondence between reciprocal-space spin splitting and real-space switching between the two quasi-altermagnetic states. These states exhibit spin splitting at $\Gamma$ point even in absence of SOC, providing a unique signature that distinguishes them from conventional altermagnetic phases. Using 2D Lieb-lattice compounds Mn$_2$WS$_4$ and its Janus derivative Mn$_2$WS$_2$Se$_2$ as representative systems, we demonstrate how modifications of atomic environment govern the emergence and evolution of distinct magnetic phases. The underlying mechanism is shown to be symmetry-driven and broadly applicable to a wide range of two-dimensional square-lattice materials. We further investigate effects of SOC, focusing on spin textures and transport signatures of the coupled quasi-altermagnetic states. We provide a sliding-induced altermagnetic and quasi-altermagnetic phases and reveal their crucial role in controlling spin degeneracy at the $\Gamma$ point, thereby paving the way for future spintronic technologies.

\section*{Conflicts of interest}
There are no conflicts to declare.

\section*{Acknowledgement}
This research was carried out with the support of the Pozna\`n Supercomputing and Networking Center (PSNC) as part of Grant No. pl0091-03 and pl0331-02. BRD (DST/INSPIRE fellowship/2020/IF200265)  highly appreciates the Department of Science and Technology (DST), Government of India, for the Inspire Fellowship. Authors also acknowledge DST FIST(SR/FST/PS-I/2022/230) for the computational facilities developed grant and the Anusandhan National Research Foundation (ANRF-SERB), Government of India, for financial support.

%\section*{References}
%\nocite{*}
\bibliographystyle{unsrt}
\bibliography{biblio}%

\end{document}